\documentclass{aastex63}
\shorttitle{Binary Parameters T CrB}
\shortauthors{Hinkle et al.}

\begin{document}

\title{
Binary Parameters for the Recurrent Nova T Coronae Borealis\footnote{Accepted Manuscript version of an 
ApJ article.  IOP Publishing Ltd is not responsible for any errors or omissions in this version of the manuscript.   
}
}

\author[0000-0002-2726-4247]{Kenneth H. Hinkle}
\affil{NSF's National Optical-Infrared Astronomy Research Laboratory,\\
P.O. Box 26732, Tucson, AZ 85726, USA}
\email{ken.hinkle@noirlab.edu}

\author[0000-0002-1386-0603]{Pranav Nagarajan}
\affil{Department of Astronomy, California Institute of Technology, 1200 East California Boulevard, Pasadena, CA 91125, USA
}

\author[0000-0002-9413-3896]{Francis C. Fekel}
\affil{Center of Excellence in Information Systems, Tennessee State University, \\
3500 John A. Merritt Boulevard, Box 9501, Nashville, TN 37209, USA}

\author[0000-0003-3457-0020]{Joanna Miko\l{}ajewska}
\affil{Nicolaus Copernicus Astronomical Centre, Polish Academy of Sciences, Bartycka 18, PL-00716 Warsaw, Poland
}

\author[0000-0002-5514-6125]{Oscar Straniero}
\affil{INAF, Osservatorio Astronomico d'Abruzzo,\\
I-64100 Teramo, Italy, and INFN, Sezione di Roma La Sapienza,  Roma, Italy}

\author[0000-0001-8455-4622]{Matthew W. Muterspaugh}
\affil{Brentwood Center and Distance Education, Los Medanos College, 
1351 Pioneer Square, Brentwood, California 94513, USA
}

\begin{abstract}
T CrB is among the brightest novae. It is recurrent with outbursts 
happening approximately every 80 years. The next outburst is imminent,
expected in 2025. The T CrB binary consists of an M4~III red giant 
(RG) secondary and a white dwarf (WD) primary.  A time series of 
spectra of the RG was obtained between 2022 and 2024.  Radial 
velocities (RVs) from these data were combined with literature RVs 
and an updated orbit computed. The orbit is circular to a high 
precision and has a period of 227.5494 $\pm$ 0.0049 days for the 
circular solution. An eccentric solution yields an eccentricity of 
0.0072 $\pm$ 0.0026. Rotational line broadening of the RG was also 
measured. Binary parameters are derived by maximum likelihood modeling
of the available observational data. The WD, in accord with other 
estimates for recurrent novae, is massive.  Assuming the 
\textit{Gaia} distance, the WD mass is $1.37 \pm 0.01\,M_{\odot}$ with 
the M giant secondary mass $0.69^{+0.02}_{-0.01}\,M_{\odot}$.
We discuss the evolution of this system and both paths to and 
limitations on further refining the values of the system parameters.
\end{abstract}

% Unified Astronomy Thesaurus concepts

\section{Introduction}\label{introduction}

T Coronae Borealis = HR 5958 = HD 143454 is a famous recurrent nova 
(RN) with outbursts traceable back
centuries \citep[][]{schaefer_2023a}.  During its 1866 eruption, it
became the first nova observed spectroscopically \citep{huggins_1866}.
The 1946 eruption was the brightest nova ever observed, peaking
at visual magnitude 2.0 \citep{schaefer_2023b}. The next nova erruption
has been predicted for 2025 $\pm$ 1.3 \citep{schaefer_2023b}.  In 
quiescence it is the brightest of the RNe with an average quiescent 
visual magnitude of 9.8 \citep{schaefer_2023b}.  
RNe are a rare class of binaries with only 11 known in the Milky Way
\citep{darnley_2021,schaefer_et_al_2022,shara_et_al_2024a}.  
T~CrB is the prototype of a RN subgroup that consists of a white dwarf
(WD) and a late-type giant. 

The T~CrB binary system is included in several taxonomies.
As a result it is discussed in papers spanning a wide 
range of topics. The quiescent light curve reveals an ellipsoidal 
variable of visual full amplitude 0.3 mag \citep{bailey_1975, 
schaefer_2023b}. Ellipsoidal variables are late-type -- WD binaries 
where the late-type star fills its Roche lobe.  The late-type star can
be a dwarf, giant, or asymptotic giant branch (AGB) star but must be 
a star currently evolving to a larger radius.  Most ellipsoidal 
variables are dwarfs since the evolutionary time for a dwarf far 
exceeds that of a giant or AGB star. As a 
Roche lobe filling system, T~CrB is a member of the wider group of 
cataclysmic variables \citep[CVs;][p. 300]{warner_1995}.
\citet{selvelli_et_al_1992},  \citet{ilkiewicz_et_al_2023}, and 
others have used similarities between T~CrB and various 
subgroups of CVs to explore the nova aspects of T~CrB.  

T CrB is also a symbiotic binary \citep[SySt;][]{merrill_1944}.  
The stars in this class are identified by spectra that in quiescence have  
cool giant absorption features combined with high 
excitation emission lines.  These systems consist of 
a red giant (RG) and a degenerate star, usually a WD, 
typically having orbital periods in the range of a year to decades.  
The hallmark of a SySt is mass exchange from the late-type 
star to the degenerate companion.  The SySt RNe form a group 
that currently includes only four known systems in the Milky Way 
\citep{anupama_mikolajewska_1999, anupama_2008}.  The small number 
of objects should not be taken to reflect the significance of
this group of objects.  The Milky Way sample is certainly not 
complete, and all novae are recurrent \citep{ford_1978}.  RNe are those
novae that have recorded re-eruptions, limiting the time between
eruptions to less than $\sim$100 years.  Over 1100 nova 
candidates have been identified in M31 \citep{pietsch_2010}, with 18 known to 
be RNe \citep{darnley_2021}.   

A RN in M31, M31N2008-12a, has an extraordinarily short
recurrence period of about one year.  
Ejecta from all nova eruptions sweep up the local interstellar
medium (ISM).  In RN ejecta overtaking previous ejecta and interacting with the ISM result in a 
nova super-remnant \citep[NSR;][]{darnley_et_al_2019,healy-kalesh_et_al_2023}.
A NSR of 134 pc diameter has been observed around M31N2008-12a \citep{darnley_et_al_2019}.  
However, other than around this exceptional RN, NSRs have proven difficult to detect, perhaps
because of intrinsic low luminosity \citep{healy-kalesh_et_al_2024b}.  Only three others have been detected.
A $\sim32$ by $10$ pc ISM cavity surrounds the RN RS Oph \citep{healy-kalesh_et_al_2024a}. A 
50 pc diameter NSR is centered on the RN 
KT Eri \citep{shara_et_al_2024a}.   A faint 30 pc diameter NSR has been detected 
around T CrB, perhaps detected only because of the relative closeness of T CrB.
\citet{shara_et_al_2024b} notes that 
the T CrB NSR provides another route to investigate the RN processes.
From the proper motion of T CrB the NSR age is 
$\sim2\times10^5$ yrs.  The luminosity of the nebula is in agreement
with the expected kinetic energy transfer into the NSR by RN eruptions with 
standard assumptions for the mass transfer, $\dot{M}~\simeq~1\,-\,3 \times 10^{-8}~M_\odot~\rm{yr}^{-1}$, and with
about half the transferred mass ejected.  

The greater significance of SySt RNe became evident from work on stellar 
evolution in the 1980s.  The mass of a carbon-oxygen (C--O) WD 
originating from the 
evolution of a low mass star is limited to $\sim$1.1~$M_\odot$
\citep{ritossa_et_al_1996}.  During the 1985 nova eruption of 
the RN RS~Oph, the bolometric magnitude was measured. With the 
assumption that novae obey the core mass -- luminosity relation 
\citep[][p. 291]{warner_1995}, the mass of the 
WD in the RS Oph binary is $\sim$1.3 $M_\odot$. This mass implies 
that the WD must be accreting mass from the RG and is approaching 
the Chandrasekhar limit. The mass of the RS~Oph WD was later 
confirmed from orbital properties by \citet{brandi_et_al_2009}.  
Similarly, the binary parameters for T~CrB show that its WD is 
massive \citep{belczynski_mikolajewska_1998}. 
A third SySt RN, V3890 Sgr, also has been shown to contain a massive WD 
\citep{mikolajewska_et_al_2021}. 

Since the work of \citet{belczynski_mikolajewska_1998},
there have been several other investigations into the T CrB binary
parameters.
\citet{stanishev_et_al_2004} used radial velocity (RV) 
observations of the H$\alpha$ emission to estimate masses for the 
system components in the T CrB binary. They found an orbital inclination, 
$i$, of 67$^{\circ}$ with component masses 
of $M_{\rm WD}\,=\,1.37\pm0.13 {M}_{\odot}$ 
and $M_{\rm RG}\,=\,1.12\pm0.23 {M}_{\odot}$. 
However, emission lines do not necessarily trace the center-of-mass 
motion of the WD and can be problematic proxies for the WD velocity. 
In the case of T CrB this resulted in the derivation of incorrect 
masses by \citet{kraft_1958}. Nonetheless, this effort also showed 
the presence of a massive WD. More recently, \citet{tatarnikova_et_al_2013} 
found similar values for the ratio of the WD to RG mass between 0.4 and 0.8.

Since WDs obey a degenerate equation of state, the radius decreases 
and the surface gravity rapidly increases as the mass increases.  
As a result, less hydrogen can be accreted before a nova event 
\citep{shara_1981}.  
Models by \citet{prialnik_kovetz_1995}, \citet{yaron_et_al_2005},
and \citet{hillman_et_al_2016} demonstrated that  
light curve characteristics of RN can be used to determine the WD mass. 
\citet{shara_et_al_2018}
related the time between the RN outbursts, the duration of 
the nova outburst, and the mass of the WD.  
\citet{shara_et_al_2018} found a mass for T CrB of  
1.32~$M_\odot$. Based on the nova light curve, \citet{hachisu_kato_2019} 
concluded that the T CrB WD mass was 1.38 $M_\odot$.  
Observational mass determinations allow the calibration of the 
theory. 

\citet{schaefer_2023b} has predicted that T~CrB will undergo a nova 
outburst in 2025.5 $\pm$ 1.3 yr. Eighty years ago observation 
techniques were limited to the optical and were basic. The 2025 
nova outburst  will present the first 
opportunity to measure the physical characteristics of the nova.  
It has been 25 years since 
\citet{belczynski_mikolajewska_1998} published a report on the system
parameters of T CrB.  In this paper we combine RVs from the literature with new
RV observations and undertake an improved orbital solution. We then 
carry out an evaluation of
the system employing a joint solution for all parameters.   
The results should be of use in calibrating observations of the nova.
We discuss the implication of the system parameters for the evolution 
of the binary.  T CrB, a member of many classes of objects, has also 
been considered as a possible Type~Ia supernova progenitor.   

%===========================

\section{Observations}\label{observations}

A program to obtain a time series of optical high-resolution spectra  
of T CrB
was begun in 2022 at Fairborn Observatory in southeast Arizona with 
the Tennessee State University 2~m Automatic Spectroscopic Telescope 
(AST) and fiber-fed echelle spectrograph \citep{eaton_williamson_2004,
eaton_williamson_2007}. The detector is a Fairchild 486 CCD that has
a 4096 $\times$ 4096 array of 15 $\mu$m pixels \citep{fekel_et_al_2013}.
There are 48 echelle orders that range in wavelength from 3800 to 8260~\AA.
The observations were obtained with a 365 $\mu$m fiber that produces
a resolution of 0.4~\AA\ or a resolving power $R~\approx~15,000$ at 6000~\AA.
Between 2022 Apr 8 and 2024 Apr 4, 46 useful spectroscopic observations 
were acquired. Unfortunately, Fairborn Observatory and its telescopes 
were shuttered in early April of 2024 resulting in a premature end to 
this observing program.

A discussion of the velocity measurement for the AST spectra can be 
found in \citet{fekel_et_al_2009}. The reference-star line list that 
was used for measurement contains 40 lines that are relatively 
unblended in M giant spectra and range in wavelength from 5000~\AA\ to 
6800~\AA. To fit individual line profiles, a rotational broadening
function was used \citep{lacy_fekel_2011, fekel_griffin_2011}.
With the same reduction technique, unpublished velocities of several
IAU radial-velocity standards observed with the 2~m AST have an
average velocity difference of $-$0.6 km~s$^{-1}$ when compared to
the results of \citet{scarfe_2010}. Thus, we have added 0.6
km~s$^{-1}$ to each of our AST velocities. 

In high-resolution spectra the T~CrB lines appear broadened. 
From the 46 AST spectra the projected rotational velocity, $v\,sin\,i$, 
is 8.7 $\pm$ 0.4 km~s$^{-1}$ where $v$ is the rotational velocity and 
$i$ is the inclination. The technique used to measure $v\,sin\,i$ is
described in \citet{willmarth_et_al_2016} and
\citet{henry_et_al_2022}. 

The angular diameter of the RG in T CrB was measured in the 
near-infrared $H$- and $K$-bands from 18 nights of archival interferometric 
observations collected with the Palomar Testbed Interferometer (PTI) 
between 2003 November and 2005 February. A uniform disk model fitted to 
the data resulted in an orbitally averaged angular diameter of 
$0.85 \pm 0.15$ mas \citep{rogge_2011}.

%===========================

\section{Orbit}\label{orbit}

\citet{adams_joy_1921} reported the earliest RV of the 
T CrB M~giant absorption lines, $-$5 km~s$^{-1}$ in 1921 July.  
Unfortunately, the date of the observation is ambiguous, and the 
spectrum was observed at low resolution with unknown precision and 
velocity systematics.  A few other RVs were reported 
during the next two decades, but all were measured from low 
resolution observations \citep[see footnotes 4 -- 8 in][]{sanford_1949}.
The first high-resolution T CrB spectra suitable for accurate RV 
measurements were recorded at Mount Wilson Observatory by 
\citet{sanford_1949} between 1946 March and 1948 May.
The spectra were observed starting 36 days after the 1946 February 
nova outburst. \citet{sanford_1949} reported that by the time of his 
observation on 1946 March 16, T~CrB was 6.5 mag fainter than at maximum
with the characteristic M giant TiO bands visible.
\citet{kraft_1958} made additional observations and published the 
first orbit. The orbit was further refined by \citet{kenyon_garcia_1986} 
and then by \citet{fekel_et_al_2000}. \citet{planquart_et_al_2025} published
an orbit from their observations with elements in agreement with the \citet{fekel_et_al_2000}
orbit. 

%===========================

%TABLE 1 -- Summary of RV observations -- "table:rv_summary"

\begin{deluxetable}{lcrll}
\tabletypesize{\normalsize}
\tablewidth{0pt}
\tablecaption{Summary of T CrB  Radial Velocity Observations}
\label{table:rv_summary}
\tablehead{\colhead{Hel. Julian}  & \colhead{Calendar} & \colhead{No.} & \colhead{Observatory/} & \colhead{Ref.\tablenotemark{a}} \\
\colhead{Dates} & \colhead{Dates} & \colhead{Obs.} &  \colhead{Telescope/} & \colhead{} \\
\colhead{($-$2400000)} & \colhead{} & \colhead{} & \colhead{Instrument}  & \colhead{}
}
\startdata
31896 -- 32696 & 1946 Mar 16 -- 1948 May 24 & 19 & Mt. Wilson 2.5m/coud\'e & S49 \\
35584 -- 35994 & 1956 Apr 20 -- 1957 Jun 4  &  6 & Mt. Wilson 2.5m/coud\'e & K58 \\
35939          & 1957 Apr 10                &  1 & Palomar 5m/coud\'e      & K58  \\ 
45037 -- 46141 & 1982 Mar 8 -- 1985 Mar 16  & 25 & Mt. Hopkins 1.5 m, Oak Ridge 1.5m & KG86  \\ 
50569 -- 51364 & 1997 Apr 30 -- 1999 Jul 4  &  9 & Kitt Peak/coud\'e feed, 2.1 m or 4 m/Phoenix  & F00 \\
55576 -- 60105 & 2011 Jan 14 -- 2023 Jun 9  & 100 & La Palma 1.2 m Mercator/HERMES & P25 \\
59678 -- 60405 & 2022 Apr 8 -- 2024 Apr 4   & 46 & Fairborn 2 m/fiber-fed echelle & Current\\
\enddata
\tablenotetext{a}{
S49 = \citet{sanford_1949},
K58 = \citet{kraft_1958},  
KG86 = \citet{kenyon_garcia_1986},
F00 = \citet{fekel_et_al_2000},
P25 = \citet{planquart_et_al_2025}
}
\end{deluxetable}

%===========================

% TABLE 1 summarizing RVs = table:rv_summary
% TABLE 2  RVs and fit = table:rv_obs

The 46 new AST velocities reported here have been combined with the 
RV measurements of \citet{sanford_1949}, \citet{kraft_1958}, 
\citet{kenyon_garcia_1986}, \citet{fekel_et_al_2000}, and 
\citet{planquart_et_al_2025}.  The RV observations are summarized in 
Table \ref{table:rv_summary}, with a total of 206 velocities spanning 
the period 1946 Mar 16 to 2024 Apr 4 (Table \ref{table:rv_obs}), 
covering a substantial portionof the interval between 
nova outbursts. For the four earlier data sets, we adopt the velocity
weights determined by \citet{fekel_et_al_2000}, where each KPNO velocity
was given unit weight. We first obtained an orbital solution for just
the AST velocities with the differential corrections program SB1  
\citep{barker_et_al_1967}. We then compared the resulting center-of-mass 
velocity and variances of the velocities with that of the KPNO 
velocity solution. 
The final three AST velocities, acquired 
at the end of March and beginning of April in 2024 have a systematic 
residual of about $-$1.9 km~s$^{-1}$, more than a 3$\sigma$ difference, 
compared to this initial AST orbit. Thus, they were given zero weight, 
and the orbit was recomputed. 
Comparing the AST and KPNO orbits, we found no significant difference 
in the two center-of-mass velocities, so no velocity shift was applied 
to the AST velocities. Comparing the reciprocal of the variances for 
the AST and KPNO solutions, we found essentially identical weights, 
and so like the KPNO velocities, the AST velocities were assigned unit 
weights.  Similar analysis resulted in a velocity shift of +0.5 km s$^{-1}$ 
and unit weights for the \citet{planquart_et_al_2025} velocities. 

An SB1 orbital solution of the seven sets of weighted velocities 
with all the elements varied produced an orbit with a very small 
eccentricity of 0.0072 $\pm$ 0.0026 and a longitude of periastron, 
$\omega$, of $222\degr\,\pm\,18\degr$. Because of the extremely low 
eccentricity, we also computed a circular orbit 
with SB1C\footnote{D. Barlow 1998, private communication}, a program that 
iterates sine/cosine fits by differential corrections. When 
the eccentric and circular orbital-element solutions are compared, the 
precepts of \citet{lucy_sweeney_1971} indicate that the circular 
solution should be adopted. 

\citet{planquart_et_al_2025} found that 
the standard deviation of the $O-C$ values, $\sigma_{\rm OC}$, for their 
velocities was 0.5 km s$^{-1}$, larger than the expected instrumental 
uncertainty of 0.07 km s$^{-1}$.  They attributed this result to 
velocity variations intrinsic to the M giant. We have reported similar 
variations in other SySt \citep[e.g.][]{fekel_et_al_2000}. 
Periodogram analysis of the $O-C$ values in both our RVs and the \citet{planquart_et_al_2025} RVs
shows variation of full amplitude 1 km s$^{-1}$ with possible periods of 450 d and 1710 d. 
Both periods are longer than those expected in field RG \citep{soszynski_et_al_2007}.

%===========================

%TABLE 2 -- RV Obs "table:rv_obs"

\begin{deluxetable}{ccrrrl}
\tabletypesize{\normalsize}
\tablewidth{0pt}
\tablecolumns{6}
\tablecaption{T CrB Radial Velocity Observations and Circular Orbit Fit\tablenotemark{a}}
\label{table:rv_obs}
\tablehead{\colhead{Hel. Julian Date}  & \colhead{Phase} &
\colhead{RV} & \colhead{$(O-C)$\tablenotemark{b}} &
\colhead{Weight} & \colhead{Source\tablenotemark{c}} \\
\colhead{HJD$-$2,400,000} & \colhead{} &
\colhead{(km~s$^{-1}$)}  & \colhead{(km~s$^{-1}$)}  & \colhead{}
& \colhead{}
}
\startdata
 31895.964 & 0.587 & $-$44.2 &    4.3  &  0.02  & S49 \\
 31925.872 & 0.719 & $-$32.4 &    0.3  &  0.02  & S49 \\
 31954.885 & 0.846 & $-$20.0 & $-$5.7  &  0.02  & S49 \\
 31957.785 & 0.859 & $-$10.4 &    2.3  &  0.02  & S49 \\
 32070.688 & 0.355 & $-$44.0 & $-$1.3  &  0.02  & S49 \\
 32195.042 & 0.901 & $-$11.6 & $-$3.3  &  0.02  & S49 \\
 32196.007 & 0.906 &  $-$6.2 &    1.8  &  0.02  & S49 \\
 32224.976 & 0.033 &  $-$3.4 &    1.0  &  0.02  & S49 \\
 32256.957 & 0.174 & $-$16.8 &    0.0  &  0.02  & S49 \\
 32258.946 & 0.182 & $-$16.6 &    1.4  &  0.02  & S49 \\
\enddata
\tablenotetext{a}{Table \ref{table:rv_obs} is published in its entirety in machine-readable format.
A portion is shown here for guidance regarding its form and content.}
\tablenotetext{b}{$O--C$ = Observed minus calculated RV}
\tablenotetext{c}{
S49 = \citet{sanford_1949} (Zero point RV shift = $-$1.4 km~s$^{-1}$),
K58 = \citet{kraft_1958},
KG86 = \citet{kenyon_garcia_1986},
F00 = \citet{fekel_et_al_2000},
P25 = \citet{planquart_et_al_2025} (Zero point RV shift = +0.5 km~s$^{-1}$),
Fair = Tennessee State University 2~m AST and fiber-fed echelle spectrograph at Fairborn Observatory
}
\end{deluxetable}

%===========================

\citet{sterne_1941} noted that small eccentricities can result from 
the elliptical distortion of the stars in ellipsoidal systems, and this
distortion produces an expected longitude of periastron of 90$^\circ$ or 
270$^\circ$.  For the elliptical SySt T~CrB, \citet{kenyon_garcia_1986} 
found an eccentricity, $e$, = 0.012 $\pm$ 0.005 and a longitude of 
periastron, $\omega$, = $80^\circ \pm 6^\circ$. Thus, they argued for the 
reality of the small eccentricity of its orbit. This work was reviewed 
and corrected by \citet{belczynski_mikolajewska_1998}, who noted that
the actual orbit is likely circular although tidal effects may make it
appear to be slightly eccentric.  The eccentricity of our orbit, 
0.0072 $\pm$ 0.0026 is 1.7 times smaller than that of 
\citet{kenyon_garcia_1986}. 
Our longitude of periastron, $222\degr\,\pm\,18\degr$, 
does not correspond to either of the expected 
elliptical distortion values. 
Given the apparently unphysical value of $\omega$, tidal effects, 
the intrinsic variability of the M giant, and referencing the precepts 
of \citet{lucy_sweeney_1971}, we feel the circular orbit 
should be the favored solution.
 
% Table 2 -- table:rv_obs
% Table 3 -- table:orbit
% Figure 1 -- fig:orbit

\begin{figure*}
\centering
\includegraphics[width=.8\linewidth]{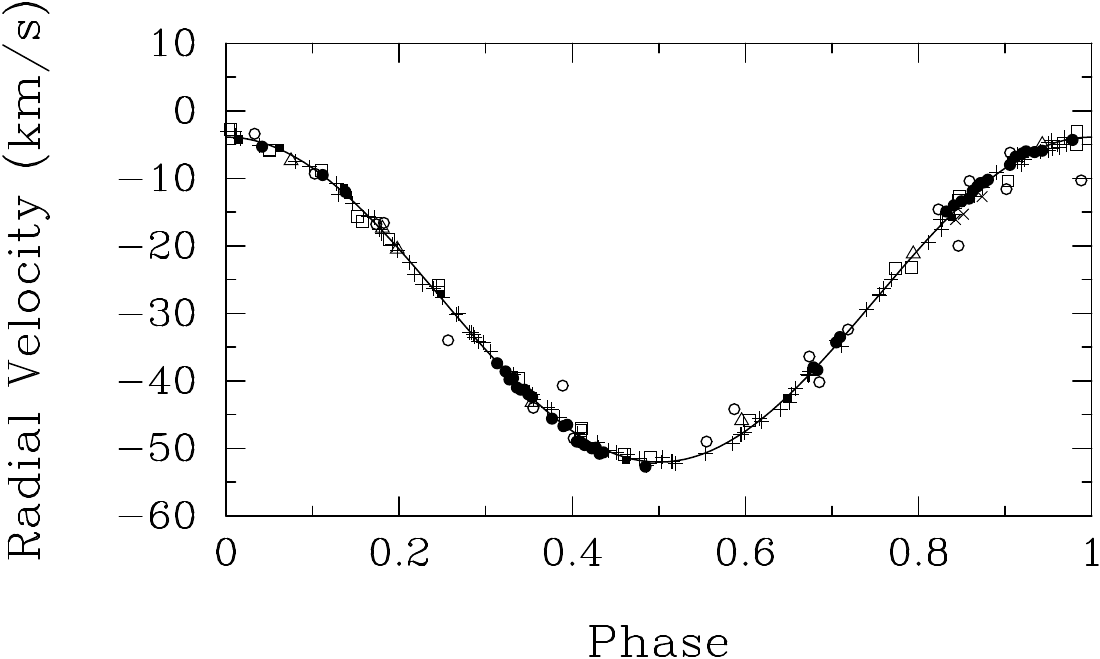}
\caption{The computed circular orbit radial-velocity curve (solid line) 
of the T CrB M giant compared with the observed velocities. Open circles =
\citet{sanford_1949}, open triangles = \citet{kraft_1958}, open squares =
\citet{kenyon_garcia_1986}, solid squares = \citet{fekel_et_al_2000},
pluses = \citet{planquart_et_al_2025}, solid circles = this paper, 
crosses = last three AST velocities (see text).
The orbital period is 227.5494 days with zero phase a time of
maximum velocity.
}
\label{fig:orbit}
\end{figure*}

The RV observations and the fit to our computed circular 
orbit are listed in Table~\ref{table:rv_obs}.
We provide both the circular and eccentric orbital elements in 
Table \ref{table:orbit}. As recommended by \citet{battenetal1989},
for a circular spectroscopic orbit, the symbol $T_0$, identified as 
a time of maximum positive velocity, is listed because there 
is no periastron passage. A comparison of the orbital elements for the
two solutions shows that for the elements in common as well as the
computed mass function, the values from the two solutions are equal to 
within their uncertainties.  Figure \ref{fig:orbit} shows the velocities 
as a function of phase for the circular orbit. In that figure the final
three AST velocities that have large negative systematic residuals fall 
between phases 0.8 and 0.9. 

%===========================

%TABLE 3 -- Orbital elements -- "table:orbit"

\begin{deluxetable}{lcc}
\tabletypesize{\normalsize}
\tablewidth{0pt}
\tablecaption{T CrB Spectroscopic Orbital Elements and Related Parameters}
\label{table:orbit}
\tablehead{\colhead{} & \colhead{Circular Orbit} & \colhead{Eccentric Orbit} \\ 
\colhead{Parameter} & \colhead{Value} & \colhead{Value}}
\startdata
$P$ (days)               & 227.5494 $\pm$ 0.0049   & 227.5464 $\pm$ 0.0049   \\
$T$ (HJD)                &  \nodata                & 2455340 $\pm$ 12  \\
$T_0$ (HJD)              & 2455427.51 $\pm$ 0.10   &  \nodata                 \\
$K$ (km s$^{-1}$)        &  24.086 $\pm$ 0.060     & 24.106 $\pm$ 0.059     \\
$e$                      &  \nodata                & 0.0072 $\pm$ 0.0026  \\
$\omega$ (deg)           &  \nodata                & 222 $\pm$ 18  \\
$\gamma$ (km~s$^{-1}$)   & $-$27.956 $\pm$ 0.042   & $-$27.978 $\pm$ 0.042  \\
$a$~sin~$i$ (10$^6$ km)  & 75.37 $\pm$ 0.19        &  75.42 $\pm$ 0.19    \\
$f(m)$ ($M_{\sun}$)      &  0.3295 $\pm$ 0.0025    & 0.3302 $\pm$ 0.0024 \\
S.E.\tablenotemark{a} (km~s$^{-1}$) & 0.5          & 0.5                     \\
\enddata
\tablenotetext{a}{Standard error of an observation of unit weight, see weights in Table 1.}
\end{deluxetable}

%===========================

% FIGURE 1  -  ORBIT

%===========================

\section{Red Giant -- Observed Parameters}\label{parameters}

A spectral type of M3 III was assigned to the T CrB RG by 
\citet{berman_1932} and confirmed by \citet{kraft_1958}.  
\citet{kenyon_garcia_1986} found that the depth of the TiO 7054~\AA\ 
feature matched that of M 3 III.
However, a detailed analysis of spectral type indicators by 
\citet{kenyon_fernandez_castro_1987} found a spectral type of M4.1 $\pm$ 
0.3 III. 

The literature contains a rich data set of optical photometry for T CrB. 
\citet{schaefer_2023b} used archived material to assemble 
the visual light curve over nearly 200 years. The nova outbursts can 
be dated much farther back. \citet{schaefer_2023a} discusses evidence 
for observations of an outburst in 1217 October.  
AAVSO visual and $V$-band data cover 
decades.  For the last few years $R$- and $I$-band AAVSO 
light curves are also available. 
Flickering, related to the mass transfer from the RG to the WD, 
becomes conspicuous in the blue.
\citet{munari_et_al_2021} noted that flickering is visible in the 
$V$-band but more so in the $U$-band. 
Similarly, \citet{walker_1977} found conspicuous flickering in $U$-band 
observations but found residual effects detectable as far red as the
$R$-band.

\citet{munari_et_al_2016} utilized the ANS Collaboration\footnote{Asiago Novae and Symbiotic stars 
Collaboration, \citet{munari_et_al_2012}} to obtain time series 
$BVR_CI_C$ data from which they constructed mean light curves. 
The contribution of the WD and/or accretion disk is conspicuous in the  
$B$- and $V$-bands and less obvious, but clearly present in the data, as  
red as the $I_C$-band.  To monitor solely the RG, near-infrared colors 
are required. Time series photometry in the $K$- and $J$-bands are available 
from \citet{tatarnikova_et_al_2013} and \citet{maslennikova_et_al_2023}. 
The \citet{maslennikova_et_al_2023} mean values for 
$J$ and $K$ are 5.91 $\pm$ 0.06 mag and 4.74 $\pm$ 0.05 mag with 
amplitudes 0.18 and 0.15 mag, respectively. So their $J-K$ is 1.17.  
Phase dependence in $J-K$, if present, is $\lesssim$0.03 mag, on the order 
of the uncertainty in the photometry.

Several papers have provided precise individual or small sets of 
near-infrared magnitudes. \citet{feast_glass_1974} reported on 
near-infrared photometry for one date, 1973 May 25, $J$ = 6.04 $\pm$ 
0.06 mag, $H$ = 5.11 $\pm$ 0.06 mag, $K$ = 4.82 $\pm$ 0.05 mag, 
$J-H$ = 0.93 mag, and $H-K$ = 0.29 mag. From seven observations 
\citet{kenyon_1988} found $K$ = 4.72 $\pm$ 0.01 mag, $J-H$ = 1.11 mag, 
and $H-K$ = 0.25 mag.
% J=6.08
\citet{munari_et_al_1992} measured $K$ = 4.78 $\pm$ 0.03 mag, $J-H$ = 0.96 
mag, and $H-K$ = 0.25 mag on 1990 June 24.
% J=5.99
The $J$, $H$, and $K$ photometry from divergent sources is in good agreement.

A visual extinction of $A_V$ = 0.2 mag was suggested as appropriate for 
the field by \citet{feast_glass_1974}, and with this reddening the colors 
are consistent with an M4 III spectral class.  
To match an M3 spectral type, additional reddening would be required.
The DUST database\footnote{
https://irsa.ipac.caltech.edu/applications/DUST/;
\citet{schlafly_finkbeiner_2011}} 
can be used to estimate the 
contribution of the interstellar medium (ISM) to the reddening.  From 
this data base,  $A_V$ = 0.18 to 0.21 mag,
consistent with \citet{feast_glass_1974}.  \citet{schaefer_2010} 
similarly found little reddening, $E_{B-V}$ = 0.1 $\pm$ 0.1 mag.  From 
Galex, $E$($B-V$) = 0.065 mag \citep{schaefer_2023b}.
The 3D dust map of \citet{green_et_al_2019} gives $E(B - V) = 0.08$ mag.
With standard interstellar reddening, $A_V$ = 0.20 mag. 
\citet{evans_et_al_2019} concluded that there 
is no detectable silicate dust in the quiescent T CrB system, hence, all the reddening is 
interstellar\footnote{Circumbinary gas ejected by the 1946 outburst has been widely discussed, 
see for example
\citet{morgan_deutsch_1947} and \citet{kraft_1958}.  The lack of dips in the T CrB post-eruption light curve
implies that the ejecta remains mainly dust free.  
Possible effects of dust in the 2025 outburst are discussed in \citet{shara_et_al_2024b} and
references therein.}.  
With the use of the standard reddening relations 
\citep{rieke_lebofsky_1985}, $A_J$ = 0.06 mag and $A_K$ = 
0.02 mag. With the latter correction for reddening, the 
\citet{feast_glass_1974} value becomes $K_0$ = $+$4.80 $\pm$ 0.02 mag.  
\citet{schaefer_2010} suggested the same value but with a slightly 
larger uncertainty, $+$4.80 $\pm$ 0.06 mag, while from 
\citet{maslennikova_et_al_2023} the value is $+$4.72 mag.
We adopt the \citet{feast_glass_1974} value for $J-K$, and so 
$J_0-K_0$ = 1.18 $\pm$ 0.03 mag. 

\citet{belczynski_mikolajewska_1998} used an effective 
temperature of 3560 K based on a spectral type of M4 III.  As discussed in 
that paper, the temperature and surface gravity must vary since the M~III 
is ellipsoidal. \citet{lines_et_al_1988} determined a prolateness coefficient 
of 0.14 $\pm$ 0.01.
A fit to the near-infrared SED by \citet{evans_et_al_2019} found $T_{\text{eff}}$ = 
3600~K, and they discussed the effects on the photosphere of the M III by 
irradiation from the WD.  Both the distortion of the RG by the Roche 
lobe and irradiation by the WD should result in variations of $T_{\text{eff}}$ 
with orbital phase. While \citet{schaefer_2023b} found the effects of 
irradiation negligible in the the $V$-band, the photometry of 
\citet{munari_et_al_2016} shows an excess 
at $V$ of $\sim$0.1 with the WD in front of the M giant.  The activity 
of the accretion disk can be a factor.  The ``super-active'' mass 
accretion event from 2015 to 2023 generated renewed focus on disk and 
M giant interaction \citep[see e.g.][]{planquart_et_al_2025}.  Detailed 
discussion of that event is beyond the scope of this paper. The 
near-infrared colors show that the orbital phase dependence of the 
effective temperature is small, and so we assume a single effective 
temperature.  

\citet{kucinskas_et_al_2005} presented relations between 
color indices and effective temperature.
The unreddened $J-K$ corresponds to $T$$_{\text{eff}}$ = 3400 $\pm$ 150~K. 
The phase dependent range in $J-K$ is less than the uncertainty.
\citet{wallerstein_et_al_2008} derived the same $T$$_{\text{eff}}$ 
from photometry 
using the calibration of \citet{alonso_et_al_1999}.  However, 
\citet{wallerstein_et_al_2008} also found $T$$_{\text{eff}}$ = 3600~K from the 
Fe~I and Ti~I spectral lines. \citet{galan_et_al_2023} used 
$T$$_{\text{eff}}$ = 3400 K. 
From the $J-K$ spectral-type calibration of \citet{tokunaga_2000}, T~CrB 
is an M4 III matching the \citet{kenyon_fernandez_castro_1987} spectral 
classification.  The \citet{tokunaga_2000}
effective temperature calibration yields 3560 $\pm$ 100 K. Using 
spectrum synthesis of the 0.75 -- 2.5 $\mu$m region, 
\citet{pavlenko_et_al_2020} set $T$$_{\text{eff}}$ = 3600 K with the 
assumption of solar abundances. 
As a compromise between photometry and spectrum synthesis, 
we assume $T$$_{\text{eff}}$ = 3500 $\pm$ 100 K.

The \textit{Gaia} distance is 887 $^{+22}_{-23}$ pc \citep{bailer_jones_2021}.
With $K_0$ = 4.80 $\pm$ 0.02 mag and the \textit{Gaia} distance,  
the absolute $K$ magnitude is $-$4.96 $\pm$ 0.07.  The 
\citet{bessell_wood_1984} bolometric correction relation at $K$ 
from $J_0-K_0$ = 1.18 $\pm$ 0.03 mag 
is 2.91 $\pm$ 0.03 mag.  Therefore,  $M_{bol}$ = $-$2.05 $\pm$ 0.09 mag.
The luminosity of the T CrB RG is 565$^{+49}_{-45}$ $L$$_\odot$.
From the luminosity and effective temperature, the radius is 
65 $\pm$ 6 $R$$_\odot$ in good agreement with 
\citet{belczynski_mikolajewska_1998} who estimated 66 $\pm$ 11 $R$$_\odot$.  

\citet{selvelli_et_al_1992} did not detect eclipses in the T CrB 
ultraviolet continuum, a stringent test for eclipses of the 
companion by the much larger RG. Similarly, eclipses have 
never been observed in optical or near-infrared colors.  This sets the 
maximum value of the rotational inclination ($i$) as a function of the 
ratio of the RG to WD mass \citep{belczynski_mikolajewska_1998}.  
\citet{zahn_1977} provided a relation for the rotational -- orbital 
synchronization of late-type stars.  For orbital periods and masses 
of the SySts, and especially shorter period systems like 
T CrB, the synchronization time scales are a few times 10$^3$ orbits 
or less.  \citet{zamanov_et_al_2007} found that the rotational and 
orbital periods of nearly all SySts, including T CrB, are 
synchronized.  With synchronization, measurement of $v\,sin\,i$ and 
the orbital period gives $i$. \citet{belczynski_mikolajewska_1998} 
adopted $i$ = 60$^\circ\pm5^\circ$ based on an upper limit for $v\,sin\,i$ 
of 10 km s$^{-1}$ from \citet{kenyon_garcia_1986}. 
\citet{tatarnikova_et_al_2013} found a similar uncertainty but a 
slightly different range with $i$ =  52$^\circ$ to 62$^\circ$. 
With $v\,sin\,i~=~8.7 \pm 0.4$ km~s$^{-1}$ 
(Section \ref{observations}) and a naive solution without correction 
for Roche overflow, specific limb darkening, etc., the orbital inclination 
is 53$^\circ$ $^{+5^\circ}_{-7^\circ}$.  The assumption, made throughout this
paper, is that the orbital and rotational axes are coaligned.

Based on near-infrared, high-resolution spectra, \citet{galan_et_al_2023} 
measured C, N, O, and Fe abundance ratios, 
finding [C/Fe] = $-$0.38
$\pm$ 0.11, [N/Fe] = +0.47 $\pm$ 0.12, [O/Fe] = $-$0.25 $\pm$ 0.10,
and [Fe/H] = +0.35 $\pm$ 0.08.  
Using optical spectra,
\citet{wallerstein_et_al_2008} found [M/H] = $-$0.1 $\pm$ 0.1.
Activity associated with the WD accretion disk produces veiling 
in the optical that is variable and can affect abundances.  This is not a problem in 
the near-infrared.
\citet{pavlenko_et_al_2020} measured isotopic ratios,  $^{12}$C/$^{13}$C = 
10 $\pm$ 2 and $^{16}$O/$^{18}$O = 41 $\pm$ 3.  While 
the $^{12}$C/$^{13}$C is a small value, it is in accord with several 
other SySt $^{12}$C/$^{13}$C values \citep{galan_et_al_2016, 
galan_et_al_2023}.  
The $^{16}$O/$^{18}$O ratio, if correct, is not explainable 
by simple evolutionary models and should be confirmed.
Overall, T CrB shows a surface composition resulting from the first dredge-up.
A high N/C ratio and a low $^{12}$C/$^{13}$C ratio are indicators of H-burning CN cycle products. 
Comparing the C, N, and O values to field first ascent giants
\citep{smith_lambert_1985, jia_et_al_2018}, the values are near the edge of
normal distributions for stars of similar temperature and luminosity. 
The CO photometric index measured by \citet{kenyon_1988} is 
0.25 $\pm$ 0.01, close to the average field M4 giant value of 0.30 
$\pm$ 0.03.  The \citet{galan_et_al_2023} [Fe/H] is at the high edge of the local
field giant distribution \citep{anders_et_al_2014}. 
The abundances determined by \citet{galan_et_al_2023} have the same 
trend away from normal as seen in the weak G band giants 
\citep{adamczak_lambert_2013}. 
T CrB is both cooler than the weak G band giants and has less extreme abundances. 

\citet{wallerstein_et_al_2008} found an overabundance of $^7$Li in the 
T CrB RG.  \citet{woodward_et_al_2020} confirmed this finding, 
measuring a lithium abundance in the M III of A(Li) = 2.4 $\pm$ 0.1. This overabundance 
can be attributed to material deposited on the surface of the M4 III 
during nova outbursts.  The time since the latest outburst is 
astronomically short, so the material has not been convectively mixed. 
\citet{munari_et_al_2021} suggested possible alternate processes.
Enhanced Li and low values of $^{12}$C/$^{13}$C are similarly seen in 
the weak G band giants.  Li abundances in the weak G band giants
suggests that for these stars there is also a 
connection to binary interaction \citep{sneden_et_al_2022}. 
In the weak G band giants rotationally induced mixing 
on the main sequence has been suggested as the root cause of the C, N, 
and O abundance anomalies \citep{adamczak_lambert_2013}. 

% TABLE 4  OBSERVED RG PARAMETERS -- table:parameters 

The observational parameters of the RG are summarized in Table \ref{table:parameters}.

%===========================

%TABLE 4 -- Observations M giant Parameters -- "table:parameters"

\begin{deluxetable}{lcl}
\tablewidth{0pt}
\tablecaption{Observationally Determined Parameters for the T CrB M Giant}
\label{table:parameters}
\tablehead{\colhead{Parameter} & \colhead{Value} & \colhead{Source}
}
\startdata
Distance              &  $896^{+22}_{-23}$ pc            & \textit{Gaia} EDR3          \\
Spectral Type         &  M4.1$\pm$0.3 III                & \citet{kenyon_fernandez_castro_1987}   \\
Effective Temperature &  $3500\pm100$ K                  & Sp.Ty., photometry   \\
Luminosity            &  $565^{+49}_{-45}~{L}_\odot$     & Distance, K$_0$ \\
Radius                & $65\pm6~{R}_\odot$               & $L$, $T$$_{\text{eff}}$ \\
Angular Diameter      & $0.85 \pm 0.15$ mas              & Orbital average, \citet{rogge_2011} \\
Inclination (i)       & $53^{+5}_{-7}$ $^\circ$             & Orbital period, $v\,sin\,i$  \\
Rotational velocity   & 10.5$^{+2.6}_{-0.7}~\rm{km~s}^{-1}$ & $v\,sin\,i$, inclination \\
\rm{[}Fe/H\rm{]}                & $+0.35\pm0.08$                   & \citet{galan_et_al_2023} \\
$^{12}$C/$^{13}$C     & $10\pm2$                         & \citet{pavlenko_et_al_2020} \\
\enddata
\end{deluxetable}

%===========================

\section{Analysis}\label{analysis}

%---------------------------------------------------------------------

\subsection{Near-Infrared Light Curve Modeling}
\label{sec:nir_lc_modeling}

If we assume a circular orbit and tidal locking, the ellipsoidal 
variation of the flux $F$ of the RG in a given filter can be described with 
a two-harmonic model,
%-----
\begin{equation}
    F = \bar{F} + \sum_{i = 1}^{2} A_i \cos \left(\frac{2 \pi i (t - T_{\text{conj}})}{P}\right),
\end{equation}
%-----
where $P$ is the orbital period, $T_{\text{conj}}$ is an epoch 
of conjunction with the RG in front of the WD, $\bar{F}$ is the mean 
flux, and $A_i$ is the variability amplitude of the $i$th harmonic. 
The log-likelihood term ($\ln\mathcal{L}$) for the light curve (LC) data, with the assumption 
of Gaussian uncertainties, including a scatter term $s^2$ to account 
for potentially underestimated errors is

%-----
\begin{equation}
    \ln{\mathcal{L}_{\text{LC}}} = -\frac{1}{2} \sum_{t} \left\{\frac{\left(F_{\text{obs}, t} - F_{\text{pred}, t}\right)^2}{\sigma_{F, t}^2 + s^2} + \ln{\left[2 \pi \left(\sigma_{F, t}^2 + s^2\right)\right]}\right\}.
\end{equation}
%-----
We apply these relations to the $J$- and $K$-band near-infrared 
photometry for the RG from \citet{maslennikova_et_al_2023}. 
The order of magnitude of the scatter term is less than or equal to the 
order of magnitude of the photometric uncertainties provided 
by \citet{maslennikova_et_al_2023}, confirming that the uncertainties 
are not underestimated. 

We simultaneously fit the ellipsoidal variation in the $J$- and 
$K$-bands by modeling the variability amplitude $A_2$. $A_2$ 
is expected to be the dominant component \citep{morris_naftilan_1993}.
To do this, the $J$- and $K$-band light curves were initially
fit separately,
with $P$, $T_{\text{conj}}$, $\bar{F}_J$, $A_{1, J}$, $A_{2, J}$, $s_J$,
$\bar{F}_K$, $A_{1, K}$, $A_{2, K}$, and $s_K$ as free parameters.
Then, in our joint fit of all available data (see Section
\ref{sec:joint_modeling}), we fix $\bar{F}$, $A_1$, and $s$ in each
filter to the best-fit values derived from this initial fit. In the
joint fit, we predict $A_2$ in each filter from a given mass ratio
$q(M_{\text{WD}}, M_*)$, Roche lobe filling factor $f(M_{\text{WD}},
M_*, R_*, P)$, and orbital inclination $i$ using the prescription
of \citet{morris_naftilan_1993}. We fix the linear limb-darkening
and gravity-darkening coefficients in each filter based on
\citet{claret_bloemen_2011}, and apply the correction factor of
\citet{gomel_et_al_2021}, which is relevant at large Roche lobe
filling factors.

% FIGURE 2 -- fig:lc_fig

\begin{figure*}
\epsscale{1.0}
\includegraphics[width=1.0\linewidth]{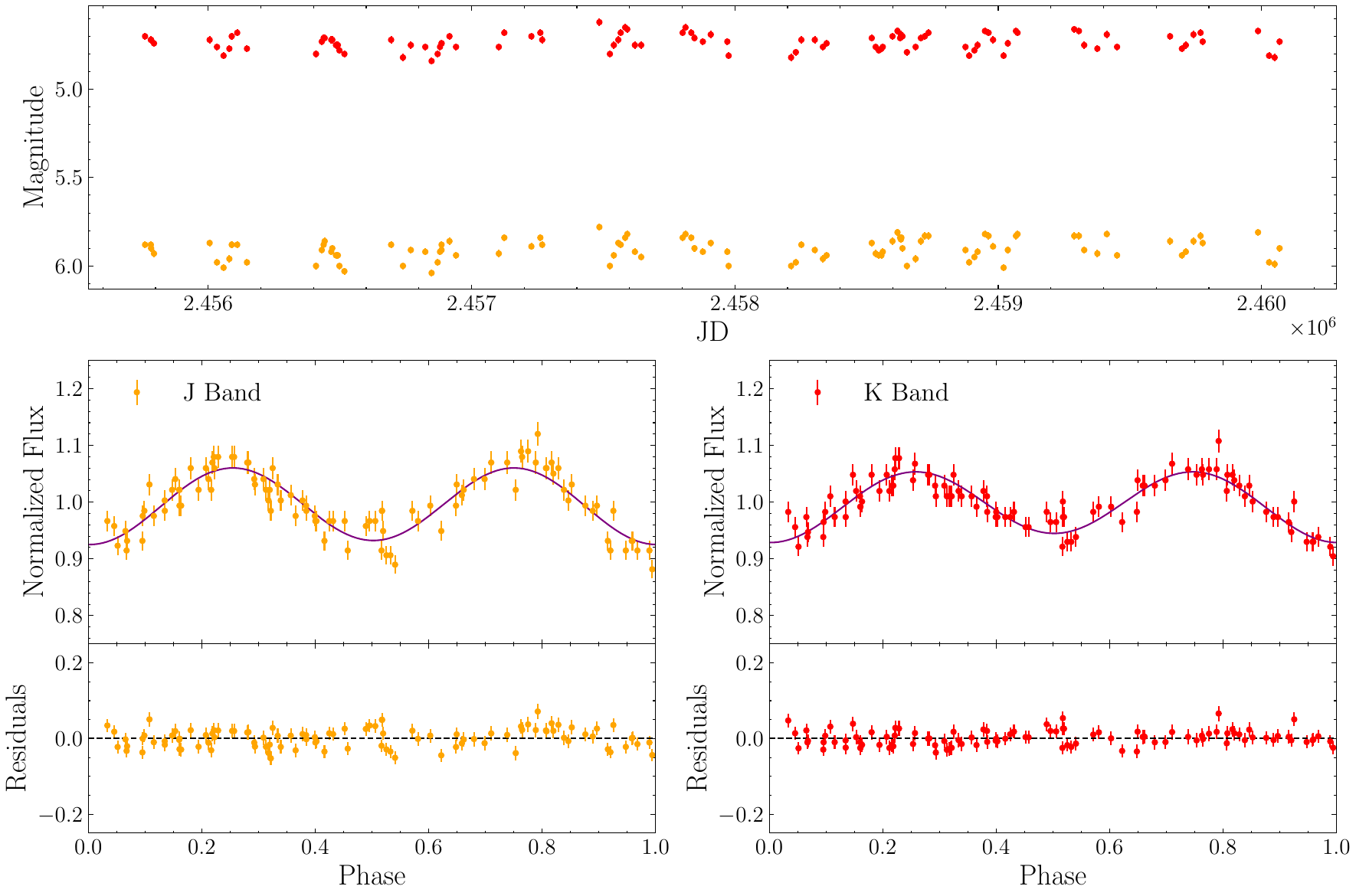}
\caption{In the top panel, we show the $J$-band (orange) and $K$-band
(red) light curves of T CrB obtained from \citet{maslennikova_et_al_2023}.
In the bottom panels, those light curves are phased with the
best-fit orbital period of $227.544$ days from the joint fit. The
purple curves represent the best-fit ellipsoidal light curve model
in each filter. In both filters, the residuals are consistent with
zero, implying that the ellipsoidal light curve model explains
almost all of the observed photometric variability in the near-infrared.}
\label{fig:lc_fig} 
\end{figure*}

The unphased $J$-and $K$-band light curves from 
\citet{maslennikova_et_al_2023} are shown in the top panel of
Figure~\ref{fig:lc_fig}. The normalized and phased light curves are
plotted for both filters in the bottom panels of Figure~\ref{fig:lc_fig}.
We overplot the best-fit ellipsoidal variability models derived
from our joint fitting of all available data (see Section
\ref{sec:joint_modeling}).

%--------------------------------------------------------------------------
% FIGURE 2 Light Curves
%---------------------------------------------------------------------

\subsection{Radial Velocity Curve Modeling}
\label{sec:rv_curve_modeling}

%--------------------------------------------------------------------
% FIGURE 3 RV Curve
\begin{figure*}
\includegraphics[width=1.0\linewidth]{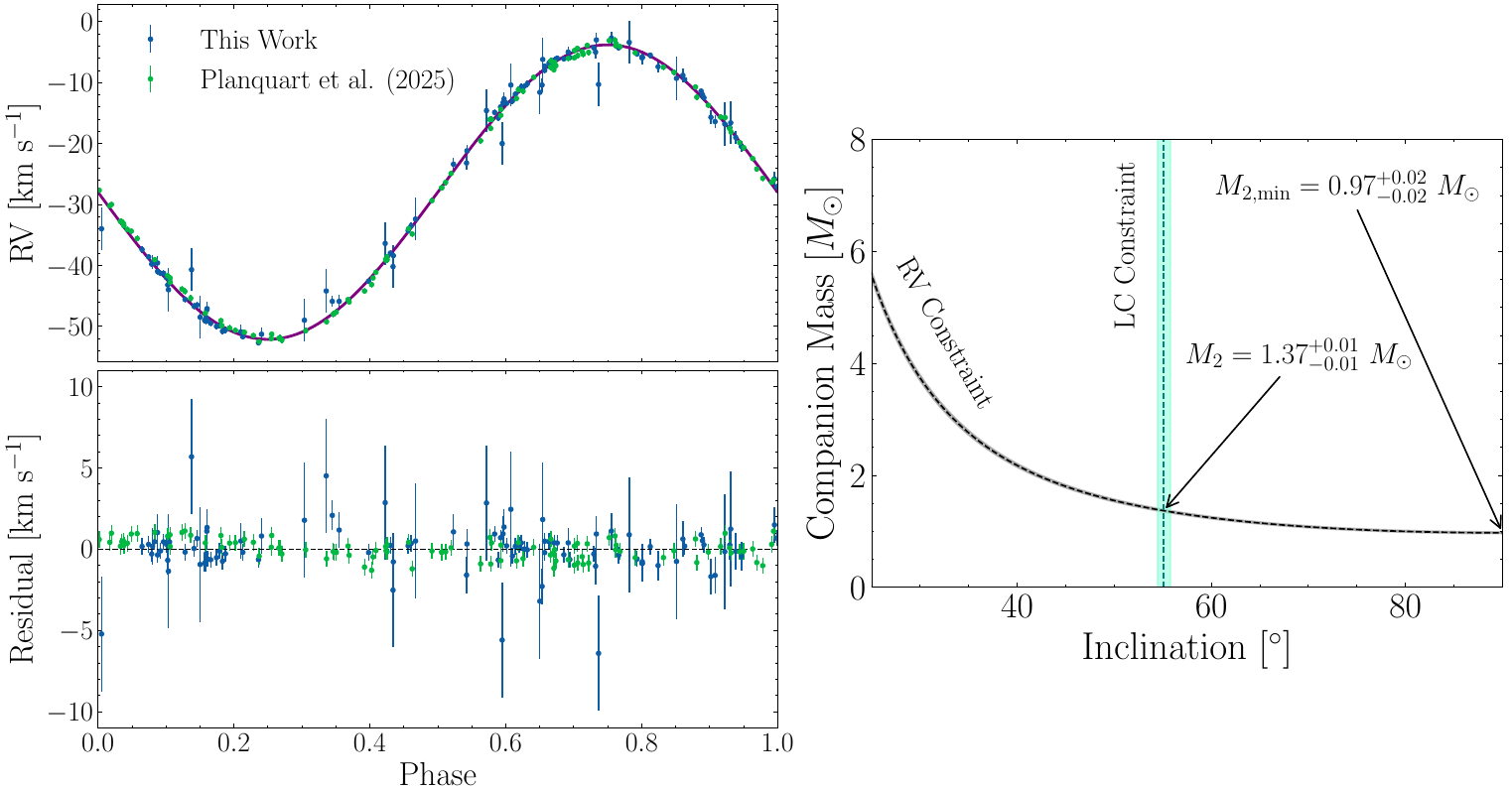}
\caption{Left Panel: RV curve and residuals of the RG in T CrB. The
RV semi-amplitude of $24.2 \pm 0.1$ km s$^{-1}$ implies a minimum
mass for the RG companion, by assuming the orbit is viewed edge on, 
of $0.97 \pm 0.02\,M_{\odot}$. Right Panel: Dynamical
constraints on the RG companion mass. When taken together, the constraints
from the measured RVs (shaded gray region) and the light curve (LC) constrains from the observed photometric
variability amplitude (shaded green region) imply $M_{\text{WD}} =
1.37 \pm 0.01\,M_{\odot}$.} \label{fig:rv_fig} 
\end{figure*}
%--------------------------------------------------------------------

%FIGURE 3 fig:rv_fig

The RVs described in Section \ref{observations} are plotted in the left 
panel of Figure \ref{fig:rv_fig}.  As in Section \ref{orbit}, we 
fit the RVs with a Keplerian two-body model,
%-----
\begin{equation}
    \text{RV}(t) = \gamma - K \sin\left(\frac{2 \pi (t - T_{\text{conj}})}{P}\right),
\end{equation}
%-----
where $\gamma$ is the center-of-mass RV, $K$ is the semi-amplitude,
and $P$ is the orbital period, and $T_{\text{conj}}$ is again 
an epoch of conjunction when the RG is in front of the WD.  
In this analysis we assume that the orbit is tidally
circularized and synchronized, consistent with the results of the
RV-only fits described in Section \ref{orbit}. The log-likelihood
term for the RV data assumes Gaussian uncertainties,
%-----
\begin{equation}
    \ln \mathcal{L}_{\text{RV}} = -\frac{1}{2} \sum_{t} \left(\frac{\text{RV}_{\text{obs}, t} - \text{RV}_{\text{pred}, t}}{\sigma_{\text{RV}, t}}\right)^2.
\end{equation}
%-----

%FIGURE fig:rv_fig

We present the best-fit RV curve from the joint fit described in
Section \ref{sec:joint_modeling}, along with its residuals, in the
left panel of Figure \ref{fig:rv_fig}. Assuming an edge-on orbit,
the observed RV semi-amplitude of $24.2 \pm 0.1$ km s$^{-1}$ implies
a minimum mass for the RG companion of $0.97 \pm 0.02\,M_{\odot}$. We plot the
inferred WD mass as a function of orbital inclination in the right
panel of Figure \ref{fig:rv_fig}. If we adopt the inclination
constraint implied by the variability amplitudes of the near-infrared
light curves (see Section \ref{sec:joint_modeling}), then we would
infer a companion mass of $1.37 \pm 0.01\,M_{\odot}$.

%---------------------------------------------------------------------

\subsection{Spectral Energy Distribution Modeling}
\label{sec:SED_modeling}

% FIGURE 4 fig:sed_fig

\begin{figure*}
\includegraphics[width=1.0\linewidth]{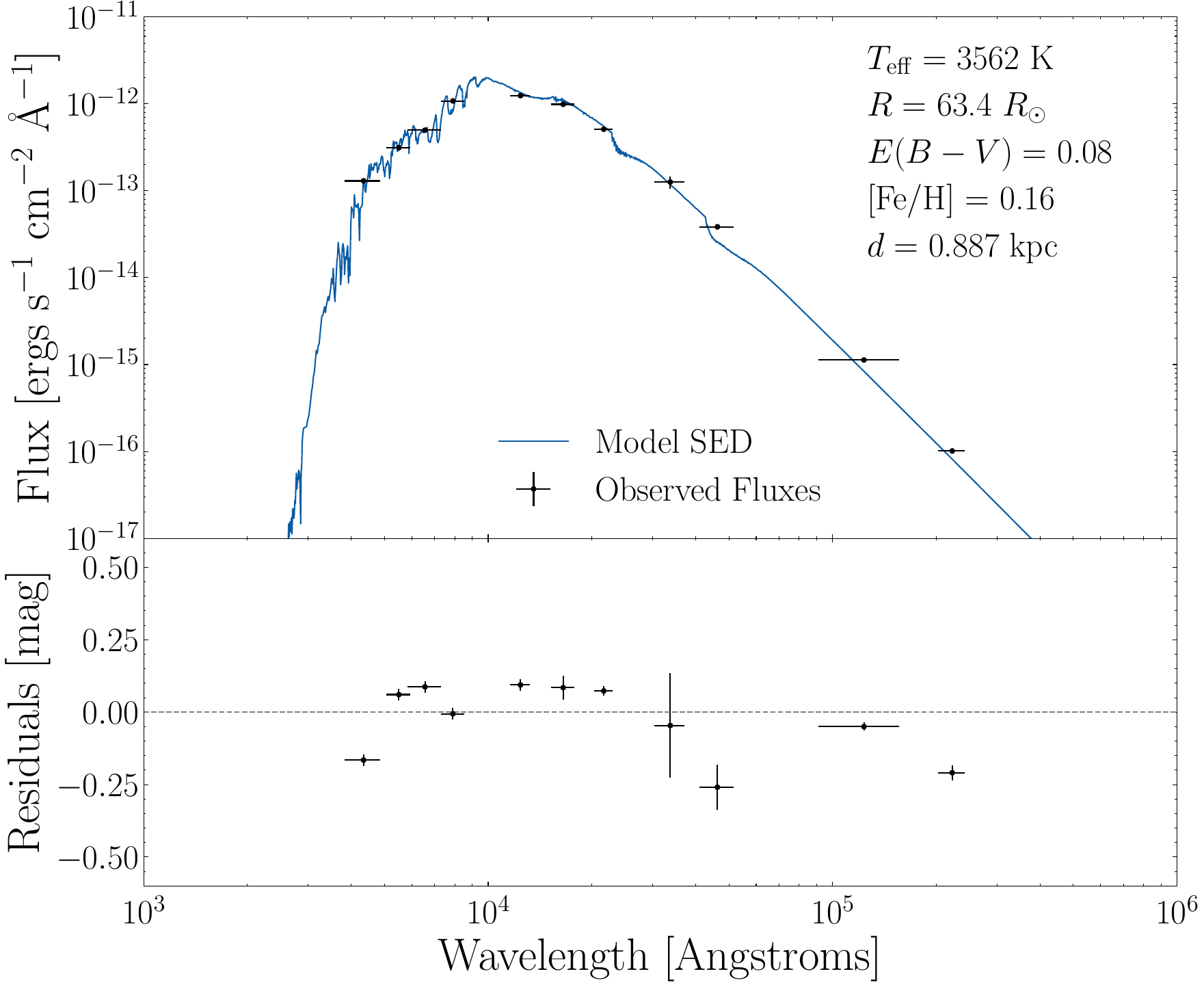}
\caption{Photometry of T CrB and modeled SED (Top) with associated residuals (Bottom)
for the RG in T CrB. In the top box,  
the vertical error bars are
derived from the uncertainties in the apparent magnitudes and the 
the horizontal bars represent the effective widths of the
corresponding bands.
The model SED is plotted based on the best-fit
stellar parameters (see Table \ref{tab:joint_fit_params}). Given
the \textit{Gaia} DR3 distance from \citet{bailer_jones_2021} and
the extinction from the 3D dust map of \citet{green_et_al_2019},
the SED analysis allows us to constrain the RG's present-day radius,
effective temperature, surface gravity, and metallicity.}
\label{fig:sed_fig} 
\end{figure*}

As discussed in Section \ref{parameters}, a time series of $BVR_cI_c$ photometry
is available from \citet{munari_et_al_2016}. T CrB entered a super-active
state in 2015, characterized by increased mass accretion on the compact object
with a corresponding increase in the hot component temperature and flux 
\citep{munari_et_al_2016}. In addition, flickering is evident in
the $B$-band light curve in both the quiescent and active states.
Hence, we use the optical photometry prior to this active period, 
from 2006--2015, to calculate
mean apparent magnitudes in the $V$, $R_c$, and $I_c$ bands, where
the contribution of the RG is dominant. We conservatively adopt an
uncertainty of $0.02$ on these mean magnitudes. We also retrieve
mean apparent magnitudes in the infrared from the 2MASS survey
\citep{skrutskie_et_al_2006} and WISE mission \citep{wright_et_al_2010}.
All of these apparent magnitudes are converted to observed fluxes
and plotted against the central wavelength of their corresponding
filters to produce an observed spectral energy distribution (SED, Figure \ref{fig:sed_fig}). 

%-------------------------------------------------------------------
% FIGURE 4 fig:sed_fig
%-------------------------------------------------------------------

% FIGURE 4

We use
\texttt{pystellibs}\footnote{https://mfouesneau.github.io/pystellibs/} with
the BaSeL\footnote{Basel Stellar Library, \citet{lastennet_et_al_2002}} library \citep{lejeune_et_al_1997, lejeune_et_al_1998}
to generate model SEDs, with the present-day mass, radius, effective
temperature, and surface metallicity of the RG as free parameters.
We fix the distance to the geometric value of 887 pc inferred
by \citet{bailer_jones_2021}. In Section \ref{sec:parallax_fit} the 
results of a joint fit where the parallax is allowed to be
a free parameter will be discussed. We assume a \citet{cardelli_et_al_1989} extinction
law with $R_V = 3.1$ and adopt a reddening of $E(B - V) = 0.08$ mag
from the 3D dust map of \citet{green_et_al_2019}, in agreement with 
values found in the literature (Section \ref{parameters}). We use
\texttt{pyphot}\footnote{https://mfouesneau.github.io/pyphot/} to
calculate synthetic photometry. The log-likelihood term for the SED
data assumes Gaussian uncertainties,
%-----
\begin{equation}
    \ln \mathcal{L}_{\text{SED}} = -\frac{1}{2} \sum_{\text{All Filters}} \left(\frac{m_{\text{obs}} - m_{\text{pred}}}{\sigma_{m}}\right)^2 .
\end{equation}
%-----
The best-fit SED model (from the joint fit described 
in Section \ref{sec:joint_modeling}), along with its residuals, is shown in Figure \ref{fig:sed_fig}. 

%---------------------------------------------------------------------

\subsection{Joint Fitting with MCMC Techniques}
\label{sec:joint_modeling}

Our joint analysis of the light curves, RVs, and
spectral energy distribution of the RG in T CrB is similar to that
described in \citet{nagarajan_et_al_2024}. However, unlike T CrB,
the SySt in \citet{nagarajan_et_al_2024} hosts
a RG donor that is not yet fully filling its Roche lobe. This enabled
the use of single-star evolutionary models to constrain the RG's
age, initial mass, and initial metallicity. In the case of T CrB,
the binary's history of ongoing mass transfer via Roche lobe overflow
renders single-star evolutionary models inapplicable; instead, only
the RG's present-day stellar parameters can be constrained.

The free parameters of our fit are the orbital period, orbital
inclination, epoch of conjunction, center-of-mass RV, and WD mass,
along with the RG's present-day mass, radius, effective temperature,
and surface metallicity. The light curves and RV curve constrain
the orbital period and epoch of conjunction. The RVs also constrain
the center-of-mass RV and the WD mass. The SED, together with the
\textit{Gaia} DR3 parallax, constrains the present-day radius,
effective temperature, and metallicity of the RG. Given the radius
and orbital period, the measured $v \sin i$ constrains the orbital
inclination. Finally, given all of the other information, the
observed ellipsoidal variability constrains the present-day mass
of the RG.

We use truncated uniform distributions to enforce that $i \in
[0^{\circ}, 90^{\circ}]$, and restrict $T_{\text{conj}}$ to fall 
within a time range spanning less than one orbital period.  We also 
use the lack of eclipses in UV photometry of T CrB to place a limit 
on the orbital inclination as a function of mass ratio \citep[see
e.g.][]{belczynski_mikolajewska_1998}. We limit the RG's present-day
surface metallicity to be less than [Fe/H] $= +0.4$ \citep[an upper
bound on the metallicity regime of RGs in the Milky Way disk, see
e.g.][]{anders_et_al_2014}. Based on theoretical simulations for
the maximum mass of a C--O WD, we place an upper limit on the mass
of the WD of $1.38\,M_{\odot}$ \citep[see e.g.][]{althaus_et_al_2023}.
The theoretical maximum mass of an oxygen--neon (O--Ne) WD is even 
lower \citep{althaus_et_al_2022}.

We jointly fit all of the available data using the total log likelihood relation,
%-----
\begin{equation}
\label{eq:log_likelihood}
    \ln \mathcal{L} = \ln \mathcal{L}_{\text{LC}} + \ln \mathcal{L}_{\text{RV}} + \ln \mathcal{L}_{\text{SED}} -\frac{1}{2}\left(\frac{v \sin i - 8.7 \text{ km s$^{-1}$}}{0.4 \text{ km s$^{-1}$}}\right)^2 -\frac{1}{2}\left(\frac{\theta - 0.85 \text{ mas}}{0.15 \text{ mas}}\right)^2,
\end{equation}
%-----
where we have added a couple of extra terms to the likelihood to
compare the RG's projected equatorial rotational velocity $v \sin
i =  2 \pi R \sin i / P$ and angular diameter $\theta = 2 R / d$
against the measured values of $8.7 \pm 0.4$ km s$^{-1}$ and $0.85
\pm 0.15$ mas, respectively.

%---------------------------------
%TABLE 5 -- Best fit parameters with Gaia -- "tab:joint_fit_params"
\begin{deluxetable}{cccc}
\tablewidth{0pt}
\tablecaption{Best Fit Orbital and Stellar Parameters with Gaia Parallax}
\label{tab:joint_fit_params}
\tablehead{
\colhead{Parameter\tablenotemark{a}} & \colhead{Description} & \colhead{Median Constraint\tablenotemark{b}} 
& \colhead{MAP Constraint}  
} 
\startdata
 \multicolumn4c{Derived from joint fit of all available data} \\
\hline 
$P$ & Orbital Period & $227.545 \pm 0.005$ days & $227.544$ days \\
	$i$ & Orbital Inclination & $55.1^{\circ\,+0.7\circ}_{-0.6^\circ}$ & $54.5^{\circ}$\\
$T_{\text{conj}}$ & Epoch of Conjunction (HJD) & $2455029.0 \pm 0.2$ & $2455029.0$\\
$\gamma$ & Center-of-Mass RV & $-27.96 \pm 0.07$ km s$^{-1}$ & $-27.97$ km s$^{-1}$ \\
$M_{\text{WD}}$ & White Dwarf Mass & $1.37 \pm 0.01\,M_{\odot}$ & $1.38\,M_{\odot}$ \\ 
$M_{\text{RG}}$ & Red Giant Mass & $0.69^{+0.02}_{-0.01} \,M_{\odot}$ & $0.68\,M_{\odot}$ \\ 
$R_{\text{RG}}$ & Red Giant Radius & $63.5 \pm 0.3\,R_{\odot}$ & $63.4\,R_{\odot}$ \\
$T_{\text{eff}, RG}$ & Red Giant Effective Temperature & $3561 \pm 3$ K & $3562$ K \\
$\rm[Fe/H]_{RG}$ & Red Giant Surface Metallicity & $+0.20^{+0.05}_{-0.03}$ & $+0.16$ \\
\hline
 \multicolumn4c{Calculated based on best-fit values from joint fit} \\
\hline
$L_{RG}$ & Red Giant Luminosity & $583 \pm 4\,L_{\odot}$ & $584\,L_{\odot}$ \\
$\log \left(g/\text{cm s$^{-2}$}\right)$ & Red Giant Surface Gravity & $0.672^{+0.009}_{-0.007}$ & $0.667$ \\
$v \sin i $ & Projected Equatorial Rotational Velocity & $11.6 \pm 0.1$ km s$^{-1}$ & $11.5$ km s$^{-1}$ \\
\enddata
\tablenotetext{a}{All stellar parameters represent present-day values.}
\tablenotetext{b}{Errors on the median constraints are derived from the 16th and 84th percentiles.}
\end{deluxetable}
%---------------------------------

% TABLE 5 - joint_fit_params
% FIGURE 5 Corner_Plot

\begin{figure*}
\includegraphics[width=1.0\linewidth]{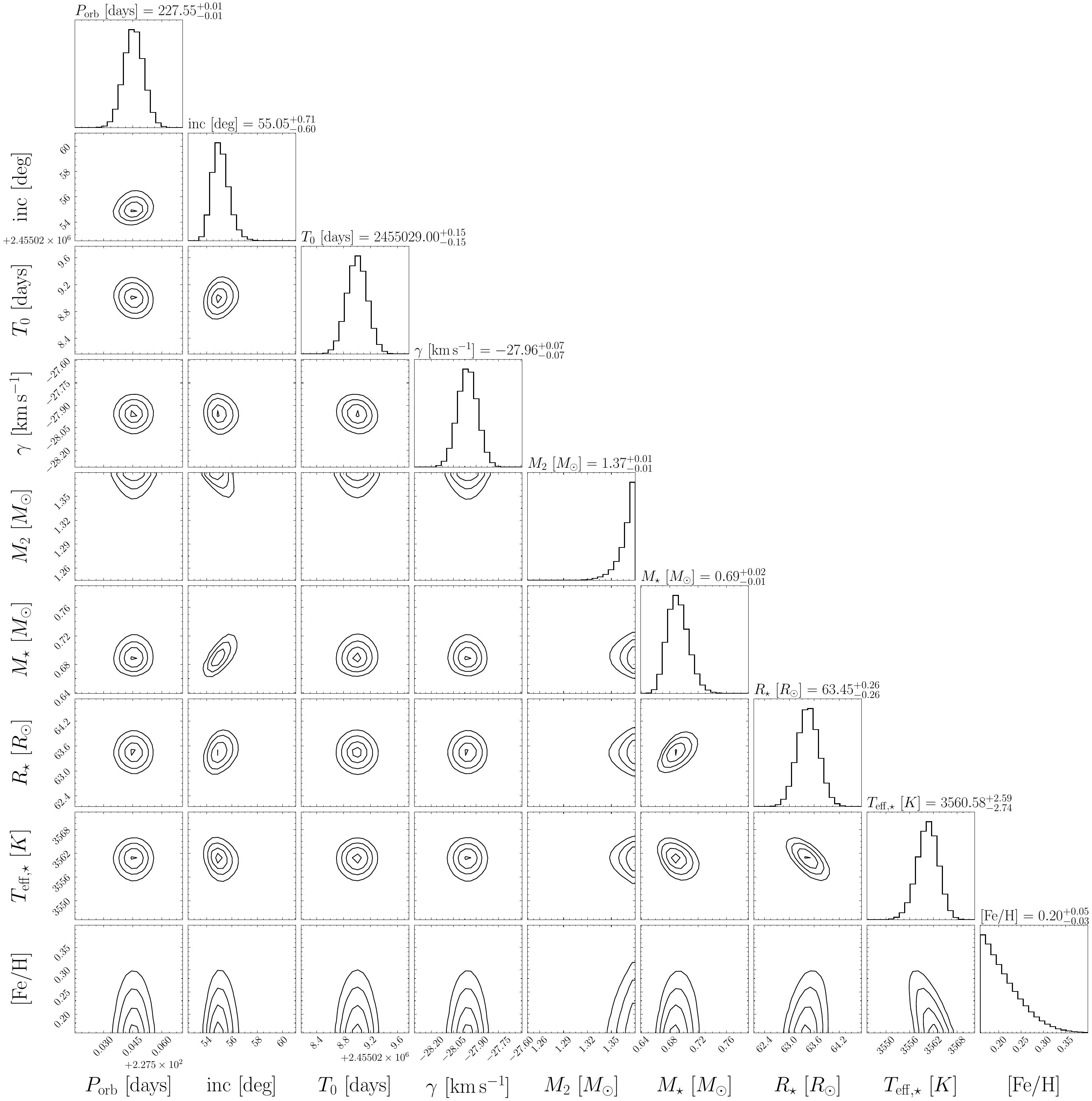}
\caption{Corner plot for orbital and stellar parameters of T CrB. 
The diagonal entries display the marginal distribution of each parameter, 
while the off-diagonal panels display the joint distributions.}
\label{fig:corner_plot}
\end{figure*}

We use ensemble Markov chain Monte Carlo (MCMC) sampling \footnote{Utilizing \texttt{\normalsize{emcee}}, \citet{emcee_2013}}
with 64 walkers and 10,000,000 total iterations to derive best-fit
orbital and stellar parameters for T CrB. We present the resulting
corner plot in Figure \ref{fig:corner_plot}, and report the median
and maximum a posteriori (MAP) constraints in Table
\ref{tab:joint_fit_params}.

%---------------------------------------------------------------------------------
% FIGURE 5 Corner plot
%---------------------------------------------------------------------------------

Our best-fit orbital parameters of $P = 227.545 \pm 0.005$ days,
$T_{\text{conj}} = 2455029.0 \pm 0.2$, and $\gamma = -27.96 \pm 0.07$ 
km s$^{-1}$ are consistent with the those from the RV-only fit 
reported in Table \ref{table:orbit}. Note again that $T_{\text{conj}}$ is 
defined as a conjunction with the RG in front, and so its phase is 
0.75 after $T_0$, which is a time of maximum velocity adopted as 
the phase zero point for the spectroscopic circular orbit.
We derive an orbital inclination of 
${i}~=~55.1^{+0.7}_{-0.6}\,^\circ$, implying that T CrB hosts a 
massive WD of $M_{\text{WD}} = 1.37 \pm 0.01 M_\odot$ that is nearing the 
Chandrasekhar limit. We derive best-fit stellar parameters of $M_{\text{RG}}  =
0.69^{+0.02}_{-0.01}\,M_{\odot}$, $R_{\text{RG}} = 63.5 \pm 0.3\,{R_\odot}$,
$T_{\text{eff}, RG} = 3561 \pm 3 \rm\,K$, and $\rm [Fe/H]_{\text{RG}} =
+0.20^{+0.05}_{-0.03}$, and confirm that the RG is Roche-lobe
filling.

\subsection{Joint Fit with Parallax as a Free Parameter}\label{sec:parallax_fit}

In Section \ref{sec:joint_modeling}, we fix the distance ($d$) to T CrB
to be 887 pc based on the zero point-corrected \textit{Gaia} DR3
parallax from its single-star solution \citep{bailer_jones_2021}.
However, the projected semi-major axis of T CrB is $\sim 0.5$ AU,
so the wobble of its photocenter on the sky has comparable contributions
from both the orbital motion and parallactic motion. In the absence
of an astrometric binary solution for T CrB \citep[which would have
a more accurate parallax with $\sim 10\times$ smaller uncertainty,
see][]{nagarajan_el_badry_2024}, the $\textit{Gaia}$ DR3 parallax
from the single-star solution still represents the best constraint
on the distance to date. However, to account for the fact that the
\textit{Gaia} DR3 parallax could be inaccurate, we also perform a
joint fit to all of the available data allowing the parallax to be a
free parameter. We use the log likelihood in Equation
\ref{eq:log_likelihood}, and additionally add a Gaussian prior on
the parallax using the zero point-corrected $\textit{Gaia}$ DR3 value
of $\varpi = 1.129 \pm 0.028$ mas. In converting to distance, we
simply take $d = 1 / \varpi$. Once again, we use ensemble MCMC
sampling \citep[\texttt{`emcee'};][]{emcee_2013} with 64 walkers and
10,000,000 total iterations to derive the best-fit orbital and
stellar parameters for T CrB. We  report the median and maximum a
posteriori (MAP) constraints in Table \ref{tab:parallax_params}.

The best-fit orbital parameters of $P = 227.546 \pm 0.005$ days,
$T_0 = 2455029.0 \pm 0.2$, and $\gamma = -27.96 \pm 0.07$ km~s$^{-1}$
are virtually unchanged between fits. However, the best-fit parallax
of $1.329^{+0.007}_{-0.006}$ mas is significantly larger than the
value reported in \textit{Gaia} DR3, translating to a significantly
smaller distance of $752^{+3}_{-4}$ pc. The modified best-fit stellar
parameters of the Roche lobe-filling RG are $M_{\text{RG}} =
0.419^{+0.002}_{-0.004}\,M_{\odot}$, $R_{\text{RG}} =
53.8^{+0.01}_{-0.02}\,R_{\odot}$, $T_{\text{eff,RG}} = 3525 \pm 3
\rm\,K$, and $\rm [Fe/H]_{\text{RG}} = +0.14^{+0.01}_{-0.03}$. Luminous, cool,
Roche lobe-filling giants of exceptionally low masses (i.e.\ $M
\approx 0.4\,M_{\odot}$) have also been observed in binaries such
as V723 Mon (``the Unicorn'') and 2M04123153+6738486 (``the Giraffe'')
\citep{el_badry_et_al_2022}. The modified best-fit projected
equatorial rotational velocity of $v \sin i = 9.1^{+0.2}_{-0.1}$
km s$^{-1}$ is in good agreement with the value of $8.7 \pm 0.4$
km s$^{-1}$ determined from optical high-resolution spectroscopy (see
Section \ref{observations}). Lastly, the best-fit orbital inclination
of $i = 50^{\circ} \pm 1^{\circ}$ implies a slightly decreased companion
mass of $M_{\text{WD}} = 1.31 \pm 0.05 {M}_\odot$.

%---------------------------------------------------------------------------------------
%% TABLE 6 -- Best fit with parallax also fit
\begin{deluxetable}{cccc}
\tablewidth{0pt}
\tablecaption{Best Fit Orbital and Stellar Parameters with Parallax a Free Parameter} 
\label{tab:parallax_params}
\tablehead{
\colhead{Parameter\tablenotemark{a}} & \colhead{Description} & \colhead{Median Constraint\tablenotemark{b}} & \colhead{MAP Constraint}
}
\startdata
\multicolumn4c{Derived from joint fit of all available data} \\
\hline
$P$ & Orbital Period & $227.546 \pm 0.005$ days & $227.548$ days \\
$i$ & Orbital Inclination & $50^{\circ} \pm 1^{\circ}$ & $50^{\circ}$\\
$T_{\text{conj}}$ & Epoch of Conjunction (HJD) & $2455029.0 \pm 0.2$ & $2455029.0$\\
$\gamma$ & Center-of-Mass RV & $-27.96 \pm 0.07 \, \rm{km~s}^{-1}$ & $-27.98 \, \rm{km~s}^{-1}$ \\
$M_{\text{WD}}$ & White Dwarf Mass & $1.31 \pm 0.05\,M_{\odot}$ & $1.31\,M_{\odot}$ \\ 
$M_{\text{RG}}$ & Red Giant Mass & $0.419^{+0.002}_{-0.004}\,M_{\odot}$ & $0.423\,M_{\odot}$ \\ 
$R_{\text{RG}}$ & Red Giant Radius & $53.8 ^{+0.01}_{-0.02} R_{\odot}$ & $54.0\,R_{\odot}$ \\
$T_{\text{eff},RG}$ & Red Giant Effective Temperature & $3525 \pm 3$ K & $3523$ K \\
$\rm[Fe/H]_{\text{RG}}$ & Red Giant Surface Metallicity & $+0.14^{+0.01}_{-0.03}$ & $+0.16$ \\
$\varpi$ & Parallax & $1.329^{+0.007}_{-0.006}$ mas & $1.324$ mas \\
\hline 
\multicolumn4c{Calculated based on best-fit values from joint fit} \\
\hline
$L_{\text{RG}}$ & Red Giant Luminosity & $403^{+2}_{-4} L_{\odot}$ & $405\,L_{\odot}$ \\
$
\log 
(
g
/
\text{cm~s}
^{-2}
) 
$
& Red Giant Surface Gravity & $0.5995^{+0.0004}_{-0.0008}$ & $0.5998$ \\
$v \sin i $ & Projected Equatorial Rotational Velocity & $9.1^{+0.2}_{-0.1}\,\rm{km~s}^{-1}$ & $9.2\,\rm{km~s}^{-1}$ \\
\enddata
\tablenotetext{a}{All stellar parameters represent present-day values.} 
\tablenotetext{b}{ Errors on the median constraints are derived from the 16th and 84th percentiles.}
\end{deluxetable}
%---------------------------------------------------------------------------------------

%===========================================================

\section{Binary Evolution Scenario}\label{binary_evolution}

Enabled by the accuracy of the stellar and orbital parameter measurements presented in 
Table \ref{tab:joint_fit_params} and \ref{tab:parallax_params},
we investigate the nature of the progenitor system of the 
RN T CrB. The general scenario is the same suggested 
by \citet{whelan_1973} for the single-degenerate path to a SN~Ia.  In
this framework, a SySt consisting of a RG donor and a
WD accretor evolves from a main sequence (MS) binary 
consisting of an intermediate-mass
primary, $3\leq M/M_\odot\leq 10$, and a low-mass companion, $M\leq2$
$M_\odot$. If the initial separation is large enough, the more
massive object evolves through H and He burnings, without
interactions with the companion, and develops a degenerate C-O
core. Then, if the C-O core mass is lower than $\sim 1.07$ $M_\odot$,
the star enters the AGB.  For core mass larger than this limit 
carbon burning takes place after which an O--Ne core forms, and the
star eventually moves to the super-AGB \citep{becker_1980,ritossa_et_al_1996}.
The minimum stellar mass for the C ignition, $M_{up}$, depends on
the metallicity. For [Fe/H]$ = 0.2$, or Z = 0.03, it is $\sim$8.3
$M_\odot$  \citep{straniero_2016}.  During the AGB phase, because
of the active burning shells, the mass of the C--O or O--Ne core
slightly increases.  A fast superwind sets in, and the
resulting loss of orbital angular momentum causes a shrinkage of
the binary until the AGB star overfills the Roche lobe.  
Because of the large mass ratio, $q = M_{\rm donor}/M_{\rm accretor}$, a
dynamical-unstable mass transfer takes place, so fast that the
companion cannot accommodate all the accreting material and a common
envelope (CE) develops. During the CE phase, the envelope of the
AGB star is lost, while the MS star is marginally affected. 
The result is a system consisting of a compact WD and a low-mass MS
star. If the initial separation was between $10^3$ $R_\odot$ and $10^4$
$R_\odot$, the post-common envelope separation of the WD$+$MS system
should be $30 < a_{f}/{R}_\odot < 150$, where a$_{f}$ is the semimajor axis 
of the WD--MS binary. Such a system is wide enough
to ensure that a second Roche-lobe overflow will occur when the MS
star will be on the red giant branch.

To reproduce the current orbital and stellar parameters
of T CrB, we simulated the post-CE evolution of the low-mass
secondary star through the MS-RG transition up to the Roche-lobe
overflow episode.  We used the binary evolution 
FuNS\footnote{FUll Network Stellar evolution code} code,
employing the version described in \citet{straniero_2020}.  
For this simulation 
we did not calculate the WD accretion explicitly but adopted
the \citet{HKNU_1999} and \citet{kato_2004} prescriptions for the growth of the
WD mass (see below). The initial MS and WD masses and
their separation were varied, searching for those systems that can
reproduce the present-day values of the following 5 parameters: RG
mass, RG luminosity and radius, WD mass and orbital period.  In
addition, the RN nature of T CrB places further constraints on the model.
This requires that the current mass transfer rate 
be less than the minimum required for stable WD accretion, but not
too small, otherwise strong H-shell flashes occur and in turn, the
efficiency of the WD mass accretion would not be sufficient to reach
its current high mass. According to \citet{shara_et_al_2018}, the accretion
rate of RNe is in the range $10^{-7}$ $M_\odot$ to $10^{-8}$
$M_\odot$. For T CrB in particular, they estimate an accretion
rate $\log \dot{M}=-7.68$. 

%---------------------------------------------------------------------------------
% FIGURE 6  accretion rate versus WD mass
\begin{figure*}
\includegraphics[width=0.8\linewidth]{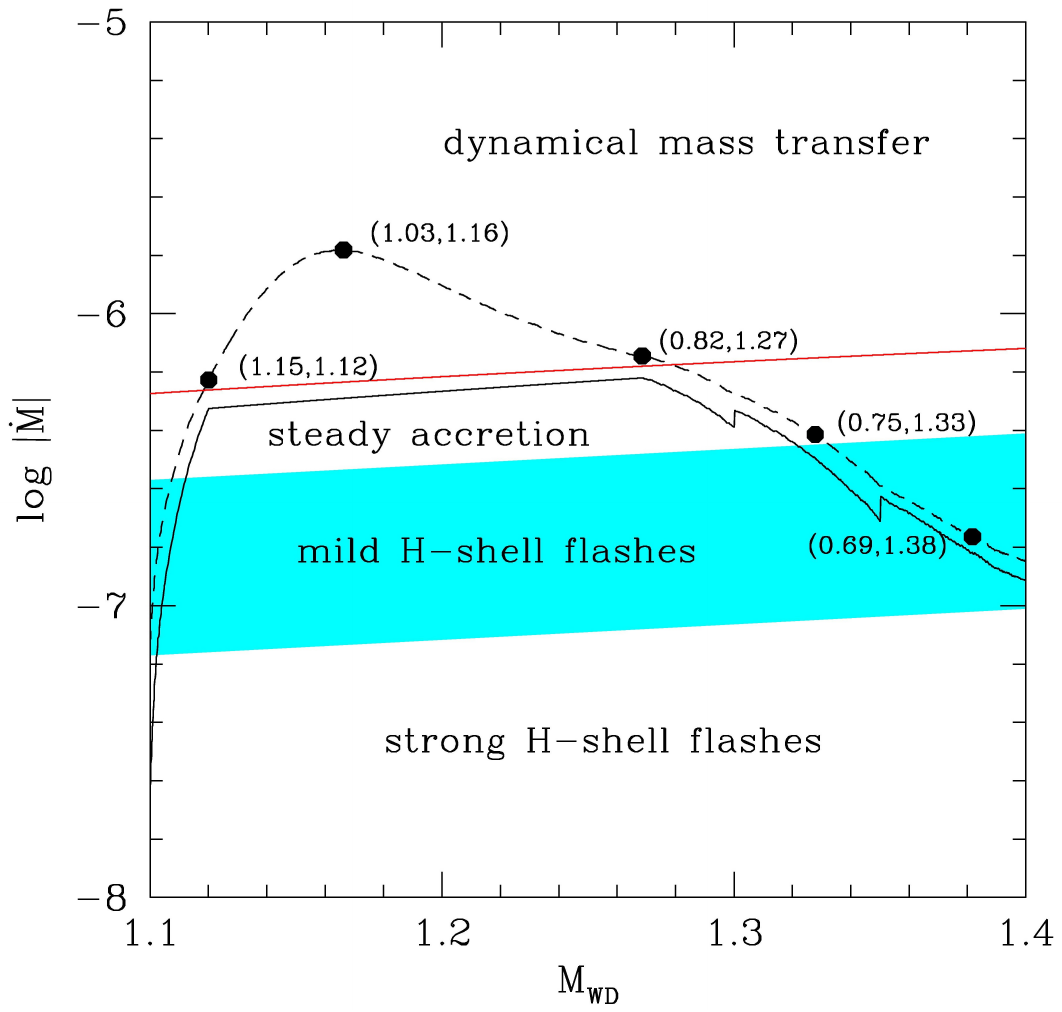}
%\resizebox{0.80\textwidth}{!}{\includegraphics{rates.pdf}}
\caption{Evolution of the mass transfer rate from the RG star
(black dashed line), and the WD mass growth rate (black solid line),
as a function of the WD mass. 
Masses are in $M_\odot$ and rates are in $M_\odot$ yr$^{-1}$. 
Dots represent key points during the mass
transfer process with corresponding masses of the two binary
components shown in brackets.  
The red line represents the
critical rate over which a dynamical mass transfer takes place 
(eq. \ref{eq:rcr}) while mild H-shell flashes occur when the accretion
rate is within the shaded area. This is also the phase where recurrent
novae are expected.}
\label{fig:rates}   
\end{figure*}
%---------------------------------------------------------------------------------

Similar to previous studies on the evolution of SySt such as
\citet{Hachisu_1999} and \citet{han_2004}, the FuNS code has been
configured for standard interacting binary evolution.  A solar
composition \citep{magg_2022} scaled to the measured metallicity,
i.e., [Fe/H]$=+0.2$ or $Z=0.03$, is adopted, while the initial
helium mass fraction is $Y=0.28$.  The mixing-length parameter,
$\alpha=1.87$, has been calibrated, as usual, by reproducing the
solar radius with a standard solar model. The initial mass of the
secondary star is varied between 1 $M_\odot$ and 2 $M_\odot$, while 
that of the WD is in the range 0.8 $M_\odot$ to 1.3 $M_\odot$.  
During the RG phase
and until the system remains detached, a Reimers mass-loss rate
%with $\eta=0.2$ 
is assumed. Then, the Roche-lobe radius and the
mass transfer rate are calculated following \citet{eggleton_2006}.
If the mass transfer rate exceeds the critial value for
stable accretion ($M_{\rm cr}$), we assume that hydrogen burns steadily
at the critical rate on the WD surface, and the unprocessed matter
is lost from the system. The critical rate from \citet{HKNU_1999} is
\begin{equation}\label{eq:rcr}
\dot{M}_{cr}=5.3\times10^7\frac{(1.7-X)}{X}(M_{\text{WD}}-0.4)~{M}_\odot\,\rm{yr}^{-1}
\end{equation} 
where $X$ is the hydrogen mass fraction  of the matter transferred
from the RG star.  The WD mass, $M_{\text{WD}}$, is in solar masses.
If the transfer rate is lower than the critical, but larger
than $0.125\dot{M}_{\rm cr}$, a steady accretion or mild H-shell flashes take
place. In that case, the prescriptions of \citet{kato_2004} for the
H and He burning retention efficiency are assumed. 
For an even lower mass transfer rate, hydrogen-shell flashes are so strong
that no more mass is accumulated on the WD. 
An investigation into the production of elements in classical nova
by \citet{starrfield_et_al_2020} also demonstrates that the 
accretion-outburst-accretion cycle results in a WD of increasing 
mass.

Since the total angular
momentum is not conserved, due to mass lost by the system and
emission of gravitational waves, variations of orbital separation
and period must be taken into account in our models \citep[see, e.g. ][]{deloore_1992}.
According to the standard assumption, the fraction of the transferred
mass that is not accreted to the compact companion is lost instantly
and an amount of angular momentum, equal to $J_{lost} = J_{\text{WD}} \Delta
M_{\rm lost}$ where $J_{\text{WD}}$ is the specific angular momentum of the
WD and $\Delta M_{\rm lost}$ the ejected mass, is removed.

%--------------------------------------------------------------------------
% TABLE 7
\begin{deluxetable}{ccc|ccccc|c}[ht!]
\tabletypesize{\normalsize}
\tablewidth{0pt}
\tablecaption{Binary Models Hosting a C--O WD}
\tablehead{
\multicolumn{3}{c|}{Initial MS$+$WD Values} & \multicolumn{5}{c|}{Values at 0.69 $M_\odot$} & \colhead{} 
\\
\hline
\colhead{$M_{\text{MS}}$} & 
\colhead{$M_{\text{WD}}^i$} & 
\multicolumn{1}{c|}{$P^i$} & 
\colhead{$\log L$} & 
\colhead{$M_{\text{WD}}$} & 
\colhead{$P$} & 
\colhead{$R_{\text{RGB}}$} & 
\multicolumn{1}{c|}{$\log \dot{M}_{\rm acc}$} & 
\colhead{Note} 
\\
\colhead{($M_\odot$)} &
\colhead{($M_\odot$)} &
\multicolumn{1}{c|}{(days)} &
\colhead{($L_\odot$)} &
\colhead{($M_\odot$)} &
\colhead{(days)} &
\colhead{($R_\odot$)} &
\multicolumn{1}{c|}{($M_\odot~{\rm yr}^{-1}$)} &
\colhead{} 
}
\startdata
   1.10 &  1.10 & 130.0 &  2.71 &  1.352 & 217.9 & 62.30 & -6.90 &     \\
   1.20 &  1.00 &  70.0 &  2.46 &  1.198 & 120.7 & 42.18 & -7.38 & (1) \\
   1.20 &  1.05 & 110.0 &  2.67 &  1.281 & 196.7 & 58.28 & -7.22 & (1) \\
   1.20 &  1.10 & 100.0 &  2.63 &  1.386 & 181.4 & 55.09 & -6.96 & (2) \\
   1.20 &  1.10 & 120.0 &  2.71 &  1.383 & 217.6 & 62.20 & -6.86 &     \\
   1.20 &  1.10 & 130.0 &  2.75 &  1.382 & 235.4 & 65.53 & -6.82 &     \\
   1.20 &  1.10 & 140.0 &  2.78 &  1.379 & 253.6 & 68.88 & -6.78 &     \\
   1.30 &  1.05 & 130.0 &  2.77 &  1.267 & 248.3 & 68.09 & -6.91 & (1) \\
   1.30 &  1.10 & 120.0 &  2.75 &  1.380 & 234.4 & 65.36 & -6.78 &     \\
   1.40 &  1.05 & 120.0 &  2.76 &  1.236 & 240.1 & 70.80 & -6.81 & (1) \\
   1.40 &  1.10 & 100.0 &  2.69 &  1.348 & 208.4 & 60.47 & -7.07 &     \\
   1.40 &  1.10 & 120.0 &  2.77 &  1.346 & 249.6 & 68.20 & -6.89 &     \\
   1.40 &  1.10 & 130.0 &  2.81 &  1.345 & 270.5 & 71.97 & -6.82 &     \\
\enddata
\tablecomments{(1) Accretion stops when $M_{\text{WD}}< 1.3$ M$_\odot$. (2)  
Explosion occurs well before the RG star reaches $\log L/L_\odot=2.76$.
}
\label{tab:CO_WD}
\end{deluxetable}
%--------------------------------------------------------------------------

% FIGURE 6 fig:rates

Among the possible progenitors of T CrB, the
most promising systems are summarized in Table \ref{tab:CO_WD}. 
All the binaries hosting WDs with initial
mass $M_{\text{WD}}^i\leq 1.05$ $M_\odot$ enter the regime of strong H-shell
flashes when the WD mass is $<1.3 M_\odot$.  The strong H-shell flashes stop 
the growth of the WD mass before values similar to those in T CrB can 
be reached. Models with $M_{\text{WD}}^i \sim 1.1$
$M_\odot$ provide a reasonable match of the present-day orbital and
stellar parameters (Table \ref{tab:joint_fit_params} and \ref{tab:parallax_params}). In 
particular, systems with initial mass of the secondary component 
1.2 $M_\odot\,\le\, M_{\text{MS}}\,\le\,1.4~{M}_\odot$
and initial orbital period $120~\rm{days}\,\le\,P^i\,\le\,150~\rm{days}$ 
reproduce all the
measured parameters of T CrB, within their errors. Figure \ref{fig:rates}
illustrates the case $M_{\rm MS} = 1.2$ $M_\odot$, $M_{\text{WD}}^i = 1.1$ 
$M_\odot$ and $P^i=130$ days.
After about 8 Gyr since the system formation, the RG star overfills
its Roche lobe. At this stage the RG luminosity is $L = 492~L_\odot$.
Suddenly, the mass transfer rate exceeds the critical value for
steady accretion (Eq.~\ref{eq:rcr}). However, because of the rather
small mass ratio, i.e., $q=M_{\text{RG}}/M_{\text{WD}}=1.09$, a stable dynamical
mass transfer should take place, even if a large portion of the
transferred mass is lost. Later on, the mass transfer rate decreases
and a steady accretion sets in. Then, after a further decrease of
the mass transfer rate, mild H-shell flashes occur at the surface
of the WD. This is the regime expected to be populated by RNe. 

Note in particular that the $M_{\text{MS}} = 1.2~{M}_\odot$, 
$M_{\text{WD}}^i = 1.1~{M}_\odot$ and $P^i=130$ day model 
reaches a WD mass of 1.37 (or 1.38)
$M_\odot$ when the mass of the RG star is 0.70 (or 0.69) $M_\odot$.
and the orbital period is 228 (or 234) days. Both the period and mass 
are in excellent agreement with the observed values of these 
parameters for the \textit{Gaia} distance. Regarding
the predicted luminosity of the RG star, i.e. $L=555~L_\odot$, it
is $\sim 5\%$ lower than the estimated value, while the predicted
radius agrees well, within the quoted error, with the observed
value. Furthermore, the high mass of the WD (1.1 $M_\odot$) implies
a primary star with initial mass very close to $M_{\rm up}$, the maximum
mass for progenitors of C--O WDs. Therefore, taking into account
the measured metallicity, we conclude that the original  binary was
a system likely consisting of a $\sim8$ $M_\odot$ primary and a
$\sim 1.2$ $M_\odot$ secondary.

In principle, more massive primary stars (8--10
$M_\odot$) can not be excluded.  These experience a CE during the 
super-AGB phase,
leaving O--Ne WD with  mass in the range 1.1 $M_\odot$ -- 1.3 $M_\odot$.
To explore this possibility, we have computed additional
models assuming $M_{\text{WD}}^i = 1.2$ $M_\odot$. A warning 
about a potential limitation of this calculation concerns the
unknown effect of the shell-C burning on the WD mass accretion. In
the following we assume that this process does not affect the
growth of the WD, while we adopt the usual prescriptions for the H
and the He burnings.  Keeping in mind this limit, a good agreement
with the observed orbital and stellar parameters of T CrB is also
found for a MS$+$WD pair with initial masses 1.8 $M_\odot$ and 1.2 $M_\odot$,
respectively, and initial period in the range 80-100 days. In this
case, however, the mass ratio is $q = 1.5$ and a very fast mass
transfer takes place when the RG star fills its Roche lobe. $\dot
M_{\text{RG}}$ can be as large as $5\times 10^{-5}$ $M_\odot$ yr$^{-1}$, and the
mass transfer is likely unstable. As a consequence, a second common
envelope episode might occur, during which the O--Ne WD could merge
with the He-rich core of the RG star.

%===========================================================

\section{DISCUSSION}\label{discussion}

The best fit parameters (Tables \ref{tab:joint_fit_params} and 
\ref{tab:parallax_params}) generally fall well inside the uncertainty 
range of the observed parameters (Table \ref{table:parameters}) and 
demonstrate the power of combined fitting to reduce greatly the 
uncertainty. For instance, the best fit $T_{\text{eff}}$ is 3561 
$\pm$ 3 K.  Literature values for $T_{\text{eff}}$ vary
from 3400 K to 3600 K (Section \ref{parameters}) with the best fit 
$T_{\text{eff}}$ nearly identical to the value derived 
from the spectral type of \citet{kenyon_fernandez_castro_1987} 
with the \citet{tokunaga_2000} spectral type -- effective temperature 
calibation.  
The best fit metallicity, $+0.20 ^{+0.05}_{-0.03}$, is slightly outside the uncertainty
of the observed value, +0.35 $\pm$ 0.08 (Table \ref{table:parameters}).  The 
lower value is more in line with the local field distribution \citep{anders_et_al_2014}.  
The projected equatorial rotational velocity, $v\,\sin\,i$, is an exception.
The value based on the \textit{Gaia} parallax, 
11.6 $\pm$ 0.1 km s$^{-1}$ (Table \ref{tab:parallax_params}),
is significantly larger than the observed value of 
8.7 $\pm$ 0.4 km s$^{-1}$ (Section \ref{observations}).  However, the 
value from allowing the parallax to be a free parameter,    
9.1$^{+0.2}_{-0.1}$ km s$^{-1}$ (Table \ref{tab:parallax_params}),
is a good agreement.  Nevertheless, this should not be taken as convincing 
evidence for either parallax.
\citet{belczynski_mikolajewska_1998} commented that the combined
effect of limb darkening and hot component radiation can reduce the
observed value of $v\,\sin\,i$ by 15--30\% with respect to the true
value.  Similarly, arguments can be mounted for both the 
\textit{Gaia} parallax and the fitted value.  T CrB is among the novae 
for which \citet{ozdonmez_et_al_2016} could not determine a 
distance.  \citet{schaefer_2010} found 900 $\pm$ 200 pc, and he reviews 
previous distance estimates, some ranging as large as 1300 pc.  The smallest 
of these, 960 $\pm$ 150 pc, was derived from the assumption that the 
RG fills its Roche lobe \citep{belczynski_mikolajewska_1998}.  The 
\citet{schaefer_2022} value based on \textit{Gaia} combined with 
other methods is 914$^{+24}_{-22}$ pc.
The WD mass found from the \textit{Gaia} value, 1.37 $\pm$ 0.01 $M_\odot$, 
is in agreement with values from the 
nova light curve, 1.32 -- 1.37 $M_\odot$ \citep{shara_et_al_2018}.  

To compute the orbit, the groups of spectra (Table \ref{table:rv_summary}) 
from different observatories were reviewed to determine the spectrograph 
offset and the weights of individual measurements (see discussion in 
\citet{fekel_et_al_2000} and Section \ref{orbit}).  
\citet{planquart_et_al_2025} performed a similar comparison between 
the \citet{fekel_et_al_2000} orbit and an orbit derived from observations
near the pending 2025 eruption.  As discussed by \citet{planquart_et_al_2025},
a super-active mass transfer phase extended from about 2015 to 2023. 
During this phase the continuum from the accretion disk made a strong 
contribution in the optical \citep{munari_et_al_2016}.
There was concern that this could affect the RVs.  Also, the irradiation
of the M giant by the accretion disk during the super-active phase 
might induce expansion of the M III. This could alter the line profiles 
and, hence, result in apparent changes in the velocities.
Neither we nor \citet{planquart_et_al_2025} observed any changes in 
the M giant velocities in this interval.  However, the last three 
velocities recorded by us, from 2024 Mar 28 to 2024 Apr 4, are 
divergent by about 2 km s$^{-1}$ and were not used.    

Our orbit is computed from velocities spanning 80 years. Long term 
period changes are a possible concern.  Most mass lost from the 
M giant in the T CrB binary is transferred to the WD accretion disk.  
However, the [O III] and [Ne III] lines show that a circumbinary 
nebula is present \citep{planquart_et_al_2025}. The majority of the 
mass loss from the system takes place during the nova eruption.  Can 
these orbital period changes be measured?  The current precision of the 
period determined from RVs is 0.002\%.  From Kepler's third law, 
the period is related to the square root of the mass, hence in 
principal, mass changes of more than $\sim 9 \times 10^{-5}$ $M_\odot$
would be detectable.  
The mass transfer rate from the M III to the disk in T CrB is believed 
variable, with a maximum value of $\sim 7 \times 10^{-8}$ $M_\odot$ 
yr$^{-1}$ \citep{luna_et_al_2018}.
The total mass accreted during the super-active phase from 2014 to 2023 is 
$\sim 2 \times 10^{-7}$ $M_\odot$  
\citep{zamanov_et_al_2023}. The pre-super-active value is $\sim 2.5 
\times 10^{-8}$ $M_\odot$ yr$^{-1}$ 
\citep{selvelli_et_al_1992,shara_et_al_2018}. 
Over the 80 years between eruptions, 
the maximum accreted mass to the disk totals roughly 2 $\times$ 10$^{-6}$
$M_\odot$.  Assuming this is all ejected during the nova, it is still less than the 
detectable mass change found above.  
In addition, the intrinsic variability of the T CrB M III makes 
detecting changes in the period  more difficult.  
Because the RV uncertainty is intrinsic, 
precision RV measurements will not improve the
uncertainty of the T CrB orbital period.  
Table \ref{table:comparison_orbits} compares our circular orbit 
solution with and without the \citet{planquart_et_al_2025} velocities.  
\citet{planquart_et_al_2025} contributes 100 RVs, approximately half 
of the total number of velocities.  However, these observations were 
over $\sim$20 orbits. 
\citet{planquart_et_al_2025} find an uncertainty in the period of 
$\pm$0.03 days.  
The change from including the \citet{planquart_et_al_2025} data is 
typically at the level of a few sigma.
The uncertainty of 0.005 days in our circular orbit results from 
observations spanning the entire time between eruptions.   

% TABLE 8

\begin{deluxetable}{lcc}
\tabletypesize{\normalsize}
\tablewidth{0pt}
\tablecaption{Circular Orbital Elements With/Without Planquart et al. Data} 
\label{table:comparison_orbits}
\tablehead{\colhead{} & \colhead{No Planquart} & \colhead{With Planquart} \\ 
\colhead{Parameter} & \colhead{Value} & \colhead{Value}}
\startdata
$P$ (days)               & 227.5464 $\pm$ 0.0049   &  227.5494 $\pm$ 0.0049  \\
$T_0$ (HJD)              & 2455427.21  $\pm$ 0.15  &  2455427.51 $\pm$ 0.10  \\ 
$K$ (km s$^{-1}$)        & 24.23 $\pm$ 0.10        &   24.086 $\pm$ 0.060    \\
$\gamma$ (km~s$^{-1}$)   & $-$27.960 $\pm$ 0.064   & $-$27.956 $\pm$ 0.042  \\ 
$a$~sin~$i$ (10$^6$ km)  & 75.80 $\pm$ 0.31        & 75.37 $\pm$ 0.19       \\ 
$f(m)$ ($M_{\sun}$)      & 0.3352 $\pm$ 0.0041     &  0.3295 $\pm$ 0.0025   \\ 
\enddata
\end{deluxetable}

The luminosity of a RG branch star depends only on the
mass of the degenerate core. The Paczy\'{n}ski relation
relates the luminosity, $L$, 
and mass of the degenerate helium core, $M_{\rm c}$ 
\citep{paczynski_1970, tuchman_et_al_1983}.   For giants of solar 
metallicity, 
$$
L ~ = ~ (M_c/0.16 )^8 
$$
where $L$ is in $L_\odot$ and $M_{\rm c}$ is in $M_\odot$
\citep{phinney_kulkarni_1994}.  With the use of  
the luminosity from either Table \ref{tab:joint_fit_params} or \ref{tab:parallax_params}, 
the core mass of T CrB is 0.34 -- 0.35 $M_\odot$.
If we use the \citet{rappaport_et_al_1995} relation for Pop. II, 
the core mass is 0.36 -- 0.38. For red giants in Roche lobe overflow, 
the stellar radius, and hence the size of the
Roche lobe, are similarly related to the core mass 
\citep{renzini_et_al_1992, phinney_kulkarni_1994}. This approach 
gives results in agreement with the above values.
The core mass compared to the current mass of the RG, 0.42 $M_\odot$ 
for the best fitting parallax 
or 0.69 $M_\odot$ from the \textit{Gaia} parallax, indicates that 
the mass remaining before the core is exposed is 0.04 $M_\odot$ to 0.35 
$M_\odot$.  

From the discussion of its binary evolution (Section \ref{binary_evolution}),
the likely post-common envelope mass of the WD was 1.1 $M_\odot$ with 
the WD so far having accreted between 0.16 to 0.28 $M_\odot$.  
\citet{althaus_et_al_2023} found that C--O WDs more
massive than 1.382 $M_\odot$ become gravitationally unstable. 
The limit for O--Ne WDs is 1.369 $M_\odot$. 
If the current WD mass is 1.37 $\pm$ 0.01 $M_\odot$ and 
with the assumption that less than half of the remaining RG envelope can be accreted and retained by the WD
\citep{hillman_et_al_2016}, 
the remaining mass to be transferred from the RG to the WD is sufficient for 
core collapse of the C--O WD to be a likely evolutionary end.  On the 
other hand, if the current WD  mass is at the low end of our estimates, 
1.26 $M_\odot$, the fate of the binary is likely a C-O 
plus He WD binary.  

%------------------------------------------------------------------------------------

\section{CONCLUSIONS}\label{conclusions}

\citet{belczynski_mikolajewska_1998} found a mass for the T CrB WD 
of 1.2 $\pm$ 0.2 $M_\odot$. Impressive progress has been made on 
measuring the parameters of the components in the T~CrB system over 
the intervening 25 years. For example, the orbital period of 227.545 days 
current is known with an uncertainty of 7 minutes.  With the use of maximum likelihood 
fitting, significant refinement of the measured values is possible. 
For a solution based on the \textit{Gaia} parallax, the mass of the 
WD is known to better than 1\% and that of the RG to 3\%. For the 
\textit{Gaia} parallax the T~CrB WD is among the most massive known 
with the $M_{\text{WD}}$ = 1.37 $\pm$ 0.01 $M_\odot$. 

The major source of uncertainty is the distance. The current 
\textit{Gaia} distance is from a single star solution. While this 
represents the best constraint on the distance to date, T~CrB is a 
binary with the wobble of its single star photocenter due to orbital 
motion that is comparable to the parallactic motion. A fit to all the 
observed T~CrB data with the distance included as a free parameter 
results in a distance of 752 $\pm$ 4 pc and a  WD mass of $1.31 \pm 
0.05$ $M_\odot$. Similarly, the RG mass is reduced from 
0.69$^{+0.02}_{-0.01}$~$M_\odot$ to 0.419$^{+0.002}_{-0.004}$~M$_\odot$.
While this distance seems at odds with other estimates of the distance, 
it can not be excluded. An astrometric binary solution to the 
\textit{Gaia} data should result in a distance with an uncertainty that
is approximately 10 times smaller than that of the single star solution 
\citep{nagarajan_el_badry_2024}. A complication is that the system's 
activity is a factor resulting in the contribution of the accretion disk 
to the image centroid. Future releases of \textit{Gaia} including an 
astrometric binary solution for T~CrB should clarify the distance and 
result in significant reduction of the uncertainties in the other 
system parameters.

The uncertainty in the distance results in uncertainty about the future 
of the system. There is no question that the WD is massive and near the 
Chandrasekhar instability limit \citep{althaus_et_al_2022}. As discussed 
in Section \ref{binary_evolution}, the WD is continuing to accrete 
mass. Other galactic RNe with giant companions have similar, albeit 
less well-determined WD masses, supporting the view that RNe with 
evolved companions are single degenerate SN~Ia candidates. Because of 
the uncertainty in the distance, the question of possible core collapse 
in the case of T~CrB remains unanswered. In addition to refining the 
distance, we encourage observers to determine the WD abundances. The 
evolution of the binary system suggests that the WD is a C--O WD 
(Section \ref{binary_evolution}).  For a C--O WD and assuming the 
\textit{Gaia} distance, the mass is within a few hundredths of a solar
mass of the Chandrasekhar limit. The stability limit for a O--Ne WD is 
lower.   

%------------------------------------------------------------------------------------

\acknowledgments
We thank the anonymous referee for very useful comments. 
We thank Professor Ulisse Munari for sending ANS photometry of T CrB.
SM plot, developed by Robert Lupton and Patricia Monger, was used
in the production of some figures. This research was facilitated by
the SIMBAD database, operated by CDS in Strasbourg, France, the
VizieR catalogue access tool, CDS, Strasbourg, France (DOI: 10.26093/cds/vizier),
and the Astrophysics Data System Abstract Service, operated by
the Smithsonian Astrophysical Observatory under NASA Cooperative
Agreement NNX16AC86A.  
This work made use of data from the
European Space Agency (ESA) mission \textit{Gaia}
(\url{https://www.cosmos.esa.int/gaia}), processed by the \textit{Gaia} Data
Processing and Analysis Consortium (DPAC). Funding for the DPAC
has been provided by national institutions, in particular the institutions
participating in the \textit{Gaia} Multilateral Agreement.  
KHH acknowledges the NOAO Office of Science and the NOIRLab RSS group for 
support of this research through the emeritus astronomy program.
NOIRLab is managed by the Association of Universities for Research 
in Astronomy (AURA) under a cooperative agreement with the National 
Science Foundation.
Astronomy at Tennessee State University was supported by the
State of Tennessee through its Centers of Excellence program.
PN acknowledges supported by NSF grant AST-2307232.
JM acknowledges support from the Polish National Science Center grant 2019/35/B/ST9/03944.

ORCID identification numbers:

FRANCIS C. FEKEL https:/orcid.org/0000-0002-9413-3896

KENNETH H. HINKLE https:/orcid.org/0000-0002-2726-4247

JOANNA MIKO\L{}AJEWSKA https:/orcid.org/0000-0003-3457-0020

MATTHEW MUTERSPAUGH https:/orcid.org/0000-0001-8455-4622

PRANAV NAGARAJAN https://orcid.org/0000-0002-1386-0603

OSCAR STRANIERO https:/orcid.org/0000-0002-5514-6125

%========================================================

%=======================================================

\end{document}